\documentclass[reprint, superscriptaddress, amsmath, amssymb, aps, nofootinbib, floatfix]{revtex4-2}
\usepackage[utf8]{inputenc}
\usepackage{amsmath}
\usepackage{graphicx}
\usepackage{tikz}
\usetikzlibrary{arrows.meta}
\tikzset{
  myarrow/.style={
    draw=green!80!black,
    line width=1.6pt,
    -{Latex[length=2.8mm,width=2mm]}
  },
  myline/.style={
    draw=green!80!black,
    line width=1.6pt
  }
}
\usepackage{appendix}
\usepackage{xcolor}
\usepackage[T1]{fontenc}
\usepackage{braket}
\usepackage{mathtools}
\begin{document}
\title{Core-Hole Excitation Dynamics of One-Dimensional Ultracold Trapped Fermions}
\date{\today}
\author{A. Becker}
\email{andre.becker@uni-hamburg.de}
\affiliation{Center for Optical Quantum Technologies, Department of Physics, University of Hamburg, 
Luruper Chaussee 149, 22761 Hamburg Germany}
\affiliation{The Hamburg Centre for Ultrafast Imaging,
University of Hamburg, Luruper Chaussee 149, 22761 Hamburg, Germany}
\author{G. M. Koutentakis}
\email{georgios.koutentakis@ist.ac.at}
\affiliation{Institute of Science and Technology Austria (ISTA), am Campus 1,\\ 3400 Klosterneuburg, Austria}
\author{P. Schmelcher}
\email{peter.schmelcher@uni-hamburg.de}
\affiliation{Center for Optical Quantum Technologies, Department of Physics, University of Hamburg, 
Luruper Chaussee 149, 22761 Hamburg Germany}
\affiliation{The Hamburg Centre for Ultrafast Imaging,
University of Hamburg, Luruper Chaussee 149, 22761 Hamburg, Germany}
\begin{abstract}
We investigate the nonequilibrium dynamics of core-hole excitations in a one-dimensional fermionic few-body system consisting of a spin-polarized Fermi bath coupled to a single heavy mobile impurity. The bath is initially prepared in a particle-hole configuration by emptying a selected bath single-particle orbital, while the impurity is displaced with respect to the center of the bath confinement potential. The quench dynamics are initialized by suddenly switching on the impurity-bath interaction. To resolve the resulting dynamics, we combine two complementary \textit{ab initio} approaches, namely the Multi-Layer Multi-Configuration Time-Dependent Hartree method for mixtures and a multi-channel Born-Oppenheimer framework. We show that the postquench response is governed by the interaction strength, impurity confinement, mass imbalance, and the location of the initially prepared hole within the Fermi sea. The density evolution and impurity center-of-mass motion reveal a competition between mixing and demixing of impurity and bath, while the von Neumann entropy demonstrates the buildup of pronounced many-body correlations. Most importantly, the occupation dynamics of the initially emptied orbital identifies deep core holes as substantially more robust against refilling than bulk or edge vacancies. Our results establish core-hole excitations as robust dynamical many-body features in trapped ultracold fermions and provide a controlled route towards probing orthogonality response, correlation buildup, and hole refilling in real time.
\end{abstract}
\maketitle
\section{Introduction}
The non-equilibrium dynamics of interacting many-body systems remains one of the central challenges in modern physics. Although equilibrium properties can often be described in terms of effective quasiparticles, such as polarons, their real-time response following a sudden perturbation provides a direct view into the build-up of correlations, entanglement, and collective rearrangement.

A prominent example is provided by core-hole excitations in atomic and molecular systems, where the sudden removal of an inner-shell electron creates a highly non-equilibrium state. The resulting localized vacancy triggers an ultrafast reorganization of the surrounding electronic cloud and can initiate prominent relaxation channels such as Auger decay and interatomic/intermolecular Coulombic decay (ICD) \cite{Auger1925,Feifel2011,Tashiro2011,Eland2010,Cederbaum1997,Hergenhahn2011,Jahnke2020}. These processes constitute a hallmark of correlated electron dynamics, since the fate of the initially created hole is not governed by an isolated single-particle process, but by the collective response of the many-body environment. 

From a complementary many-body perspective, the sudden creation of a localized hole is also closely related to the core-hole problem known from condensed-matter physics, where x-ray absorption or emission induces an abrupt change of the local scattering potential experienced by a Fermi sea \cite{Mahan1967,Anderson1967,Nozieres1969,Ohtaka1990}. In that context, the response is governed by collective particle-hole excitations and may lead to Anderson's orthogonality catastrophe, namely the vanishing overlap between the many-body states before and after the perturbation \cite{Anderson1967}. Although the microscopic realization differs between molecular and solid-state systems, both settings raise the same fundamental question of how a many-body environment reorganizes after the sudden appearance of a localized hole.

Ultracold atomic systems provide a highly controllable alternative setting in which impurity dynamics, orthogonality-catastrophe physics, and nonequilibrium many-body response can be explored in real time. In particular, impurity quenches in fermionic environments have been proposed as a route to observe orthogonality-catastrophe phenomena in the time domain \cite{Goold2011,Knap2012,Sindona2013}, while interferometric experiments have demonstrated decoherence and ultrafast many-body dynamics of impurities immersed in Fermi seas \cite{Cetina2015,Cetina2016}. More broadly, these developments connect naturally to the physics of quantum impurities and Fermi polarons in ultracold gases \cite{Massignan2014,Schmidt2018}, as well as to the wider setting of trapped one-dimensional few-body mixtures \cite{Sowinski2019}.

Motivated by these developments, we investigate the dynamics following core-hole excitations in a fermionic few-body system consisting of a spin-polarized Fermi bath coupled to a single mobile impurity. In contrast to the reference impurity-quench settings, we prepare the bath in a particle-hole configuration by emptying a selected single-particle orbital. At time $t=0$, the interaction between bath and impurity is switched on suddenly. Hence, we mimic the abrupt appearance of a local scattering potential as in the x-ray-edge problem and drive the system far from equilibrium. This protocol serves two complementary purposes: first, it realizes a controlled analogue of the core-hole setup in a trapped ultracold-atom setting. Second, it allows us to directly monitor how the initially empty orbital is dynamically refilled through impurity-induced many-body processes. In this way, observables such as the buildup of impurity-bath entanglement, and the occupation of the emptied orbital provide a unified characterization of the resulting nonequilibrium response. To address the involved interplay of fermionic statistics, correlations and non-equilibrium dynamics, we apply robust time-dependent \textit{ab initio} approaches, given by the Multi-Layer time-dependent Hartree method for fermions (ML-X) and a multi-channel Born-Oppenheimer approach (MCBO). 

Finally, we focus on a physically relevant heavy-impurity regime, consisting of a light fermionic bath and a substantially heavier impurity subject to harmonic confinement. Such a setting is directly motivated by experimentally realized mass-imbalanced mixtures, including $^{6}$Li-$^{40}$K, $^{6}$Li-$^{53}$Cr, $^{6}$Li-$^{84}$Sr, and $^{6}$Li-$^{87}$Rb, which span mass ratios \(m_I/m_B \approx 6.7\text{--}14.5\) and allow for tunable interspecies interactions \cite{Taglieber2008,Tiecke2010,Ciamei2022b,Ciamei2022a,Silber2005,Deh2008,Ye2020}. This parameter regime is particularly well suited for our study, since the large mass imbalance enhances the separation of impurity and bath time scales and thereby provides a natural setting for analyzing quench-induced core-hole dynamics, impurity localization, bath rearrangement and the emergence of nonadiabatic many-body effects.

The structure of this work is as follows. In Sec.~\ref{sec:theory}, we introduce the underlying Hamiltonian together with the numerically exact \textit{ab initio} methods that provide the framework for the subsequent analysis. In Sec.~\ref{sec:results}, we investigate the dynamics in terms of the one-body densities, the center-of-mass position of the impurity, and many-body measures such as the von Neumann entropy and the hole occupation. Finally, we conclude with a summary and outlook in Sec.~\ref{sec:conlcusion_and_outlook}. Technical details of the employed methods are provided in Appendices~\ref{app:MLX} and \ref{app:pes_diagnostics_second_quantization}.

\section{Setup, Methodology and Observables}
\label{sec:theory}
\begin{figure}
    \centering
    \includegraphics[width=1.\linewidth]{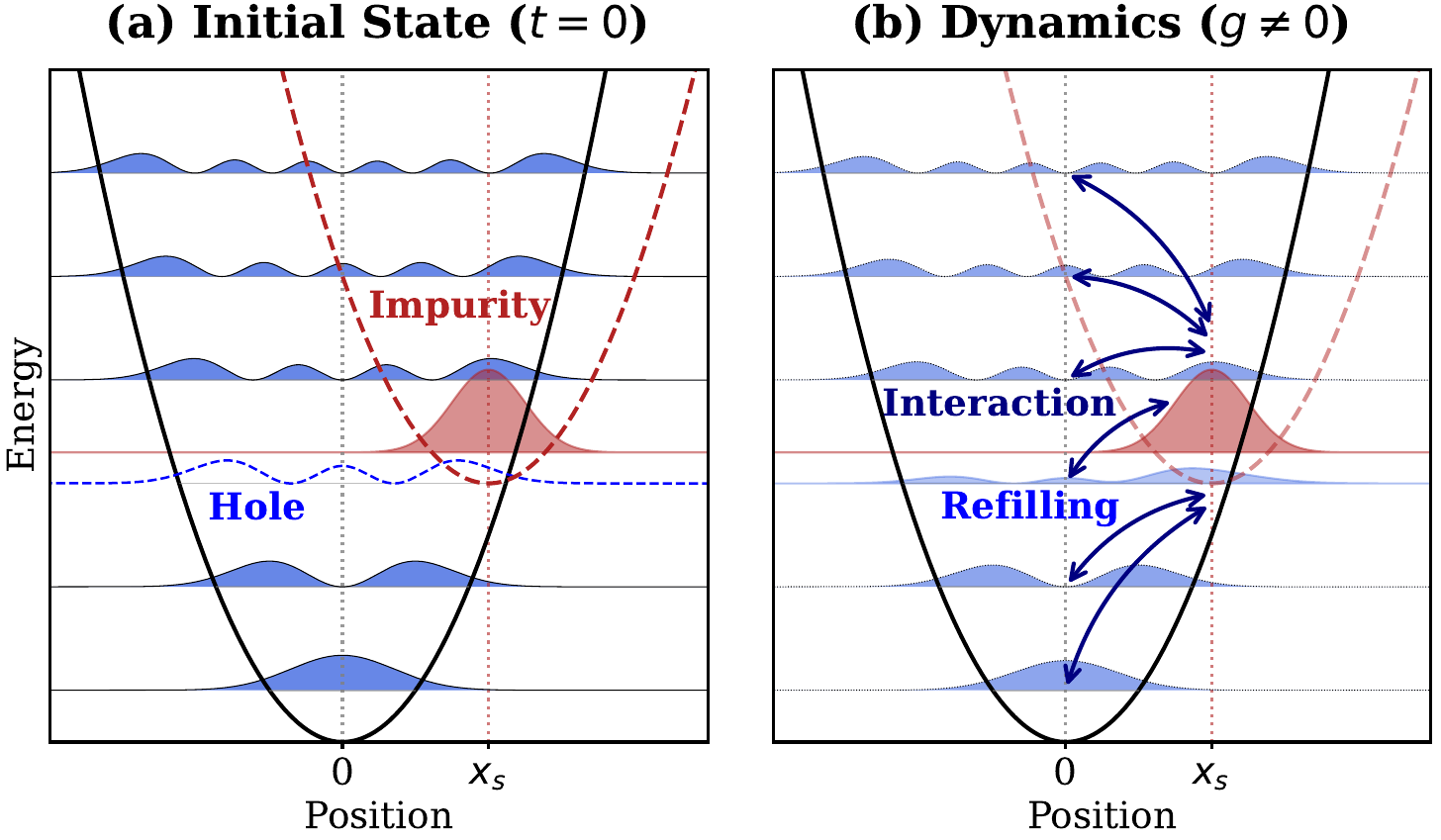}
    \caption{Schematic representation of our setup. (a) Initial non-interacting configuration of the Fermi bath with \(N_B=5\) particles and a hole excitation \(h\), together with a single displaced impurity (\(N_I=1\)). (b) Dynamics following the interaction quench $g\neq 0$, during which the bath atoms interact with the impurity, resulting in population redistribution and a refilling of the initially empty hole state.}
    \label{fig:setup}
\end{figure}
In this section, we introduce our quench protocol and Hamiltonian, summarize the two \textit{ab initio} approaches used to obtain the real-time dynamics and outline our main observables for tracking the quench-induced dynamics of the core-hole excited system.

\subsection{Quench protocol and preparation of a particle--hole excitation}
\label{subsec:quench_protocol}

We consider a bath of $N_B=5$ spin-polarized fermions and a single impurity ($N_I=1$) confined in a one-dimensional harmonic trap. We prepare both species without interaction ($g=0$), so that both bath and impurity are initially uncorrelated. The bath species follows the single-particle Hamiltonian
\begin{equation}
 \hat{H}_{B}=\sum_{j=1}^{N_B}\left[ -\frac{\hbar^2}{2m_B}\left(\frac{\partial}{\partial x_j^B}\right)^2+ \frac{1}{2}m_B \omega_B^2 \left( x_j^B \right)^2 \right],
 \label{eqn:Hamilt_bath}
\end{equation}
with $m_B$ and $\omega_B$ being the mass and trap frequency of the bath respectively.
The bath is initialized in a particular particle-hole configuration for the fixed particle number $N_B$, i.e., a vacancy (``hole'') is created in a chosen single-particle orbital, while the next free orbital is occupied instead. 
Denoting by $\ket{\mathrm{FS}(N_B)}$ the reference Fermi-sea configuration of the non-interacting bath, given by the Slater determinant obtained by filling the lowest-in-energy $N_B$ orbitals, a single particle--hole excitation reads
\begin{equation}
\ket{\Psi_B(0)} =\ket{\Psi_{B;h}} = \hat c_{p}^{\dagger}\hat c_{h}\,\ket{\mathrm{FS}(N_B)},
\label{eq:initial_ph_state}
\end{equation}
where $h \in \{0, 1, \dots, N_B-1\}$ labels the initially unoccupied (hole) orbital and $p = N_B$ the additionally occupied orbital. Hereby, $\hat c^\dagger_n$ ($\hat c_n$) creates (annihilates) a bath fermion in the
$n$-th single-particle orbital corresponding to the noninteracting bath Hamiltonian $\hat H_B$.
These operators are defined in terms of the fermionic field operators as $\hat{c}^{\dagger}_n = \int{\rm d}x~\varphi_n(x) \hat{\Psi}_{B}^{\dagger}(x)$, where $\varphi_n(x)$ is the $n$-th single-particle eigenstate of the harmonic oscillator.
They obey canonical anticommutation relations
$\{\hat c_m,\hat c_n^\dagger\}=\delta_{mn}$ and $\{\hat c_m,\hat c_n\}=0$, owing to the orthonormality of the harmonic-oscillator eigenbasis.
In this formalism, the reference is the $N_B$-body Fermi sea
$\ket{\mathrm{FS}(N_B)}=\prod_{n=0}^{N_B-1}\hat c_n^\dagger\ket{0}$.
This setup can be realized in state-of-the-art experimental realisations: single-atom controlled quantum-gas-microscopes and optical-tweezer platforms provide the key ingredients for the preparation and manipulation of tailored few-fermion configurations, including such defect-like excitations~\cite{Weitenberg2011,Haller2015, Cheuk2015, Endres2016}.

The impurity follows the Hamiltonian
\begin{equation}
 \hat{H}_{I}= -\frac{\hbar^2}{2m_I}\left(\frac{\partial}{\partial x^I}\right)^2+ \frac{1}{2} m_I \omega_I^2 \left( x^I - x_s \right)^2,
 \label{eqn:Hamilt_imp}
\end{equation}
with $m_I$, $\omega_I$ being the mass and trap frequency of the impurity respectively.
The initial state of the impurity is the ground state of a shifted harmonic trap centered at $x_s$, $\varphi^{(0)}_I(x_I)$.

At $t=0$ we perform a quantum quench by suddenly switching the interaction between the bath and the impurity to a finite value, thus driving the system far out of equilibrium \cite{Olshanii1998, Bergeman2003}. Since we are considering a fermionic setup, intraspecies interaction in the bath species is excluded due to the Pauli principle. This protocol mimics the sudden appearance of a scattering potential in the bath (core-hole problem), while enabling direct access to the ensuing real-time refilling dynamics and correlation buildup \cite{Ohtaka1990}.

The total Hamiltonian of the interacting system reads $\hat{H} = \hat{H}_B + \hat{H}_I + \hat{H}_{BI}$, where 
\begin{equation}
 \hat{H}_{BI}=\sum_{k=1}^{N_B} g\delta(x_k^B-x^I), 
 \label{eqn:Hamilt_Full}
\end{equation}
where $g$ is the interaction strength connected to the $s$-wave scattering length and transverse confinement according to \cite{Olshanii1998}. The interaction can be tuned via a magnetic field exploiting Feshbach resonances or via the transverse confinement by utilizing confinement induced resonances \cite{ChinGrimm}.

As pointed out in the introduction, we focus on a heavy impurity in a weaker trapping confinement, $\omega_I<\omega_B$, which was identified in our previous works Refs.~\cite{Becker2024, Becker2025} as a favorable regime for pronounced quench-induced dynamics and is therefore particularly suitable for analyzing the accompanying nonadiabatic effects. This directly leads us to the setup defined by $m_I=5.0\,m_B$ and $\omega_I=0.5\,\omega_B$, which we shall refer to in the following as the reference setup. A sketch of this setup is shown in Fig.~\ref{fig:setup}. Motivated by this substantial mass difference between the impurity and the bath species, we employ the MCBO approach. 

In the following, we use harmonic oscillator units based on $m_B$, $\omega_B$ and $\hbar$. Thus, the length scales are expressed in terms of $\ell_B = \sqrt{\frac{\hbar}{m_B \omega_B}}$. Within this system, the time unit is $\omega_B^{-1}$ and the interaction unit is $\hbar \omega_B \ell_B$.

The sudden interaction quench drives the initially uncorrelated impurity-bath system into a correlated nonequilibrium regime. To describe this dynamics quantitatively, we employ two complementary \textit{ab initio} approaches. We first introduce the numerically exact ML-X method, which provides the full many-body benchmark description, and subsequently turn to the MCBO framework, which is motivated by the heavy impurity species.

\subsection{The ML-X method}
The Multi-Layer Multi-Configuration Time-Dependent Hartree method for mixtures (ML-X) is a variational, \textit{ab initio}, and numerically exact approach to simulate the non-equilibrium quantum dynamics of bosonic and fermionic particles, as well as their mixtures \cite{KroenkeCao2013, CaoLushuai2013, CaoBolsinger2017}. It employs a multi-layered ansatz that variationally optimizes the quantum basis across multiple structural levels of the total many-body wavefunction. This expansion adapts dynamically to inter-particle correlations at the single-particle, single-species, and multi-species levels, enabling efficient and accurate simulations.

The total wavefunction, 
\begin{equation}
|\Psi(t)\rangle = \sum_{k=1}^D \sqrt{\lambda_k(t)} |\tilde{\Psi}_k^B(t)\rangle |\tilde{\Psi}_k^I(t)\rangle,
\end{equation}
is represented using a Schmidt decomposition of rank \( D \). This decomposition expands the many-body wavefunction in terms of single-species functions \( |\tilde{\Psi}_k^{\sigma}(t)\rangle \), where \( \sigma \) denotes the species. The Schmidt weights, \( \lambda_k(t) \), quantify the entanglement between different species, providing a compact representation of the system dynamics.
Each single species function \( |\tilde{\Psi}_k^{\sigma}(t)\rangle \) is expanded in terms of Fock states made up of \( d^{\sigma} \) time-dependent single-particle functions (SPFs), \( |\phi_i^{\sigma}(t)\rangle \), which dynamically adapt to the evolving system. These SPFs are further expressed on a discrete variable representation (DVR) basis, enabling an efficient and precise representation of the spatial degrees of freedom.
A detailed derivation of this multi-layer representation, including the role of the hierarchical structure of the ansatz, can be found in Appendix~\ref{app:MLX}.

Ground states are determined via imaginary-time propagation. The Hilbert space truncation is characterized by the configuration $C = (D; d^B; d^I)$, chosen to adequately account for interspecies entanglement and bath state occupation.

While ML-X provides a numerically exact treatment of the full correlated dynamics, the pronounced mass imbalance of the underlying system also suggests a complementary reduced description in terms of impurity motion on bath-induced potential-energy curves, which will be introduced in the following.

\subsection{The multi-channel Born-Oppenheimer approach}
\label{mcBO_main_text}

Motivated by the presence of a heavier and thus less mobile impurity, it is natural to complement the full many-body treatment by a Born-Oppenheimer-type description~\cite{BornOppenheimer1927}. To this end, we employ a generalized MCBO ansatz
\begin{equation}
\begin{split}
\Psi(x^B_1,\dots,x^B_{N_B}, x_I ; t) &= \\ \sum_{j = 1}^M \Psi_{j,I}(x_I ; t) 
&\underbrace{\Psi_{j,B}(x^B_1,\dots,x^B_{N_B};x_I)}_{\equiv\langle x^B_1,\dots,x^B_{N_B} | \Psi_{j,B}(x_I) \rangle}.
\end{split}
\label{eqn:multi-channel_BornOppenheimer}
\end{equation}
Here, \(|\Psi_{j,B}(x_I)\rangle\) denotes an orthonormal bath basis with a parametric dependence on the position of the impurity \(x_I\), where \(j=1,2,\dots\). The impurity wavefunctions \(\Psi_{j,I}(x_I;t)\) are the expansion coefficients in the many-body basis of the coupled system and obey the normalization condition
\begin{equation}
\sum_{j=1}^{M}\int dx_I\, |\Psi_{j,I}(x_I;t)|^2 = 1.
\end{equation} 

Applying the Dirac-Frenkel variational principle \cite{Dirac1930annihilation,frenkel1934wave},
\begin{equation}
\langle \delta \Psi | \hat{H} - i\hbar \partial_t | \Psi \rangle = 0,
\end{equation}
with the ansatz of Eq.~\eqref{eqn:multi-channel_BornOppenheimer}, we can derive the time-dependent equation of motion
\begin{equation}
    \begin{split}
i\hbar \frac{\mathrm d}{\mathrm d t}\Psi_{k,I}(x_I,t) &= - \frac{\hbar^2}{2 m_I} \sum_{j,l = 1}^M \left( \delta_{kj} \frac{\partial }{\partial x_{I}} -i A_{kj}(x_I) \right) \\
& \hspace{0.2cm} \times \left( \delta_{jl} \frac{\partial }{\partial x_{I}} -i A_{jl}(x_I) \right) \Psi_{l,I}(x_I) \\
& + \sum_{l = 1}^M \bigg( \delta_{k l} \varepsilon_k (x_I) + \delta_{kl} \frac{1}{2} m_I \omega^2_I (x_I-x_s)^2 \\
&  \hspace{0.2cm} + V^{\rm ren}_{kl}(x_I) \bigg) \Psi_{l,I}(x_I, t).
\label{eqn:schroedinger-time-dendent}
    \end{split}
\end{equation} 
Here, \(A_{kj}(x_I)= i\langle \Psi_{k,B}(x_I)|\partial_{x_I}\Psi_{j,B}(x_I)\rangle\) denotes the nonadiabatic derivative coupling, which measures how strongly the bath eigenstates change with the position of the impurity due to the impurity-bath interaction.

A natural and practical choice for the bath basis is given by the eigenstates of \(\hat H_B+\hat H_{BI}\) for the fixed impurity position \(x_I\), namely
\begin{equation}
\langle \Psi_{k,B}(x_I)|\hat H_B+\hat H_{BI}|\Psi_{l,B}(x_I)\rangle
=
\delta_{kl}\,\varepsilon_k(x_I),
\label{eqn:potential_energy_surfaces}
\end{equation}
with \(\varepsilon_k(x_I)\) the potential energy curves (PECs). These curves define the adiabatic energy landscape through which the impurity moves. In addition to the PECs, Eq.~\eqref{eqn:schroedinger-time-dendent} contains the derivative couplings \(A_{kj}(x_I)\), which mediate interchannel transfer, and a geometric renormalization term \(V^{\rm ren}_{kl}(x_I)\). Explicit derivations and further diagnostics are given in the Appendix~\ref{app:pes_diagnostics_second_quantization}. For details on the derivation of the corresponding static equations, see~\cite{Becker2024}.

Before proceeding, we specify the implementation of the interaction quench. Although real-time propagation is performed with the interacting Hamiltonian at \(g\neq 0\), the initial state is prepared in the noninteracting case \(g=0\). Hence, the initial state must be projected onto the postquench MCBO basis.

At \(t=0\), the total state is factored into the ground-state impurity wavefunction \(\varphi_I^{(0)}(x_I)\) and the bath state \(|\Psi_B(0)\rangle\) of Eq.~\eqref{eq:initial_ph_state}, which contains the particle-hole excitation. However, for \(t\ge 0\), the MCBO expansion employs the interacting bath eigenstates \(|\Psi_{j,B}(x_I)\rangle\) associated with \(\hat H_B+\hat H_{BI}(g)\) at fixed impurity position \(x_I\). Therefore, the initial channel amplitudes are obtained as
\begin{equation}
\Psi_{j,I}(x_I;0)=\varphi_I^{(0)}(x_I)\,\braket{\Psi_{j,B}(x_I)|\Psi_B(0)}.
\label{eq:mcbo_initial_projection}
\end{equation}
This projection determines the initial populations of the adiabatic and nonadiabatic PECs at the quench and thereby sets the stage for subsequent real-time dynamics.

After having specified both the quench protocol and the theoretical methods used for the time evolution, we now introduce the observables that will be used to characterize the resulting real-space dynamics, correlation buildup, and hole refilling.
\subsection{Dynamical observables}
\label{subsec:observables}

To characterize the postquench dynamics, we monitor real-space one-body densities, the impurity center-of-mass position as an indicator of mixing and demixing dynamics, the impurity-bath entanglement entropy, and the occupation of the initially emptied bath orbital.

The initial particle-hole excitation is prepared in the noninteracting bath basis. Accordingly, the hole index \(h\) labels a harmonic-oscillator orbital of the uncoupled bath Hamiltonian. For the present setup with \(N_B=5\) fermions, we focus on three representative hole indices: \(h=0\), corresponding to a core hole deep inside the Fermi sea; \(h=2\), corresponding to a bulk hole; and \(h=4\), corresponding to a hole at the Fermi edge, i.e., close to the Fermi surface. As discussed in the Introduction, this distinction is physically important because the energetic constraints and Pauli blocking depend strongly on the location of the vacancy within the Fermi sea and therefore have a pronounced impact on the subsequent dynamics.

We define the local density operators in terms of the field operators \(\hat{\Psi}_\sigma(x)\), as
\begin{equation}
\hat{n}_\sigma(x)=\hat{\Psi}_\sigma^\dagger(x)\hat{\Psi}_\sigma(x),
\qquad
\sigma\in\{B,I\}.
\end{equation}
The corresponding one-body densities are
\begin{equation}
\rho_\sigma^{(1)}(x;t)=\bra{\Psi(t)}\hat{n}_\sigma(x)\ket{\Psi(t)}.
\label{eq:rho1_general}
\end{equation}
They provide a direct real-space measure of the postquench response. Within ML-X, the one-body densities are straightforwardly obtained from the reduced one-body density operators, while in the MCBO picture they can additionally be resolved into impurity-channel contributions \cite{Becker2024, Becker2025}, which allow a determination of the non-adiabaticity. In particular, by calculating $\Delta\rho_\sigma^{(1)}(x;t)=\rho_\sigma^{(1)}(x;t)-\rho_\sigma^{(1)}(x;0)$, we can characterize the spatial evolution of the one-body density of species $\sigma$. We further calculate the impurity center-of-mass position
\begin{equation}
\langle x_I(t)\rangle
=
\int d x\, x\, \rho_I^{(1)}(x;t),
\label{eq:xI_mean}
\end{equation}
which tracks the impurity trajectory and provides a compact diagnostic of interaction-induced mixing and demixing between the impurity and the fermionic bath. Because the bath is centered at $x=0$ and the impurity is initially displaced, impurity motion toward the center corresponds to mixing, while motion away from the bath cloud corresponds to demixing. This can be understood as the impurity accessing states of lower or higher energy, which are associated with tendencies toward component miscibility and immiscibility, respectively.

To quantify the buildup of impurity-bath correlations, we employ the Schmidt decomposition of the total many-body wavefunction as provided naturally within ML-X,
\begin{equation}
|\Psi(t)\rangle=\sum_{k=1}^{D}\sqrt{\lambda_k(t)}\,
|\tilde{\Psi}_k^B(t)\rangle |\tilde{\Psi}_k^I(t)\rangle,
\end{equation}
where \(\lambda_k(t)\) are the natural populations of the Schmidt decomposition. From them, we compute the von Neumann entropy
\begin{equation}
S_{\mathrm{vN}}(t)
=
-\sum_{k=1}^{D}\lambda_k(t)\ln \lambda_k(t).
\label{eq:SvN}
\end{equation}
The growth of \(S_{\mathrm{vN}}(t)\) signals the buildup of genuine impurity-bath entanglement. Since the Schmidt decomposition is intrinsic to ML-X, this quantity is particularly direct to evaluate in that framework. In the MCBO analysis, the natural populations are obtained by diagonalizing the reduced density matrix constructed from the channel amplitudes in Eq.~\eqref{eqn:multi-channel_BornOppenheimer} and the corresponding bath-state overlap kernel, rather than from a Schmidt decomposition directly. The agreement with the ML-X results therefore serves as an additional indicator of convergence.

A basis-resolved measure of the refilling dynamics is provided by the occupation of the initially emptied bath orbital,
\begin{equation}
n_h(t)=\bra{\Psi(t)}\hat{c}_h^\dagger \hat{c}_h\ket{\Psi(t)},
\label{eq:nh_def}
\end{equation}
where \(\hat{c}_h^\dagger\) and \(\hat{c}_h\) create and annihilate a bath fermion in the noninteracting orbital \(\varphi_h(x)\), respectively. 
By construction, \(n_h(0)=0\) describes the empty hole state. Because \(n_h(t)\) refers to the occupation of the specific, initially emptied noninteracting orbital, it directly measures the persistence or refilling of the prepared core ($h=0$), bulk ($h=2$), or edge-hole ($h=4$) excitation. Consequently, it is the most direct observable for identifying whether a dynamically robust core-hole excitation is realized after the quench. 

In the following, these observables are used to analyze the aspects of the nonequilibrium evolution: the one-body densities provide a real-space view of the postquench response, \(\langle x_I(t)\rangle\) tracks mixing and demixing, \(S_{\mathrm{vN}}(t)\) quantifies impurity--bath entanglement,  and \(n_h(t)\) directly measures the dynamical stability of the prepared core-, bulk-, and edge-hole excitations.
\section{Results}
\label{sec:results}

We now turn to the nonequilibrium dynamics following the sudden switch-on of the impurity--bath interaction at \(t=0\). Our main goal is to identify how the initially prepared particle--hole excitation affects the subsequent real-space and many-body dynamics, and in particular under which conditions a robust core-hole excitation can be dynamically maintained. The results presented in the following are obtained using the MCBO method for computational efficiency. 
To ensure their reliability, we benchmark against ML-X calculations, which provide an \textit{ab initio} numerical benchmark. 
In particular, the von Neumann entropy constitutes a stringent test of convergence; a detailed analysis is provided in Appendix~\ref{app:Convergence}.

\subsection{Real-space density dynamics: core, bulk, and edge-hole response}
\label{subsec:results_density}
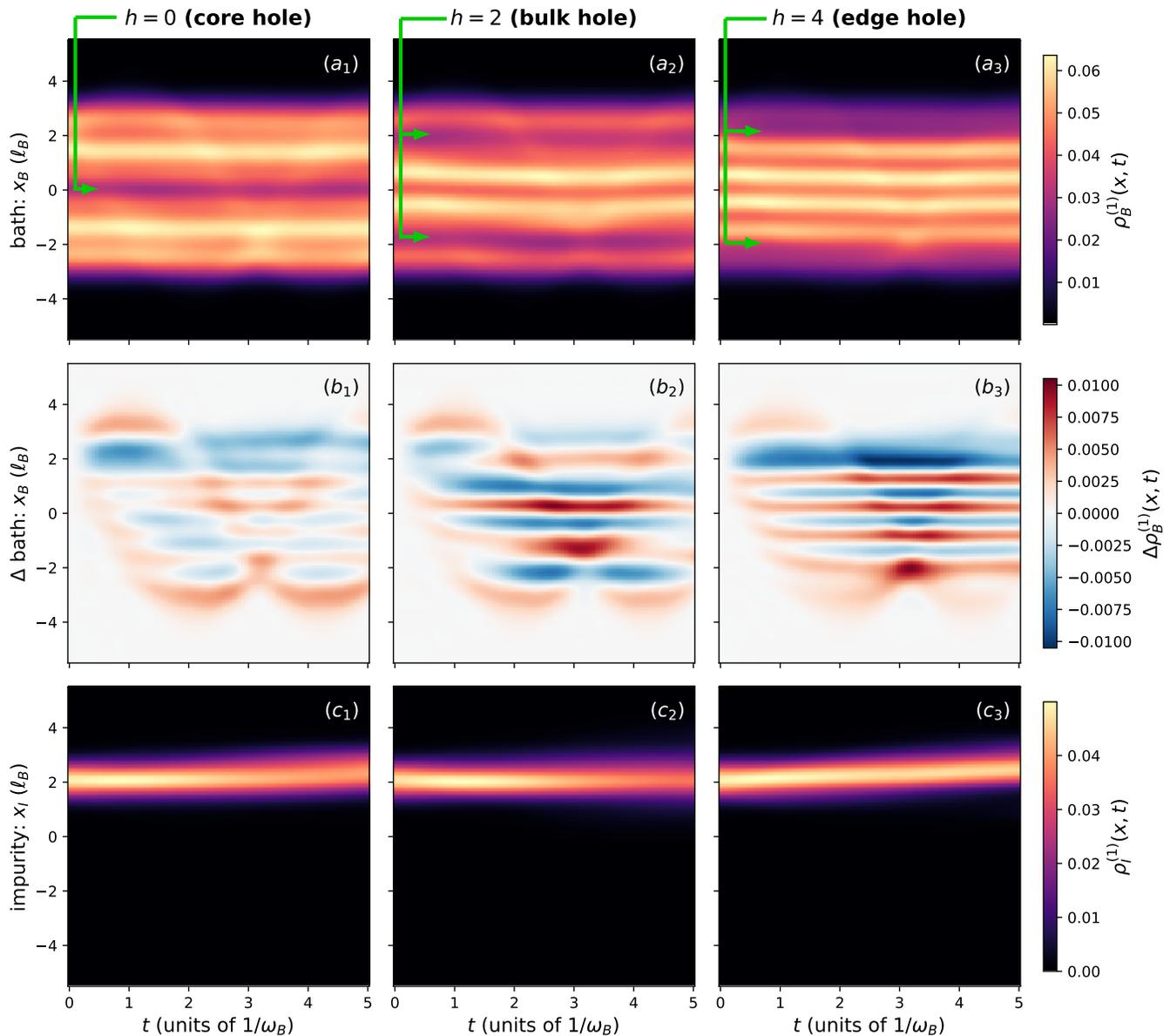
\begin{figure*}[t]
    \centering
    \begin{tikzpicture}
  \node[anchor=south west, inner sep=0] (img) at (0,0)
    {\includegraphics[width=\linewidth]{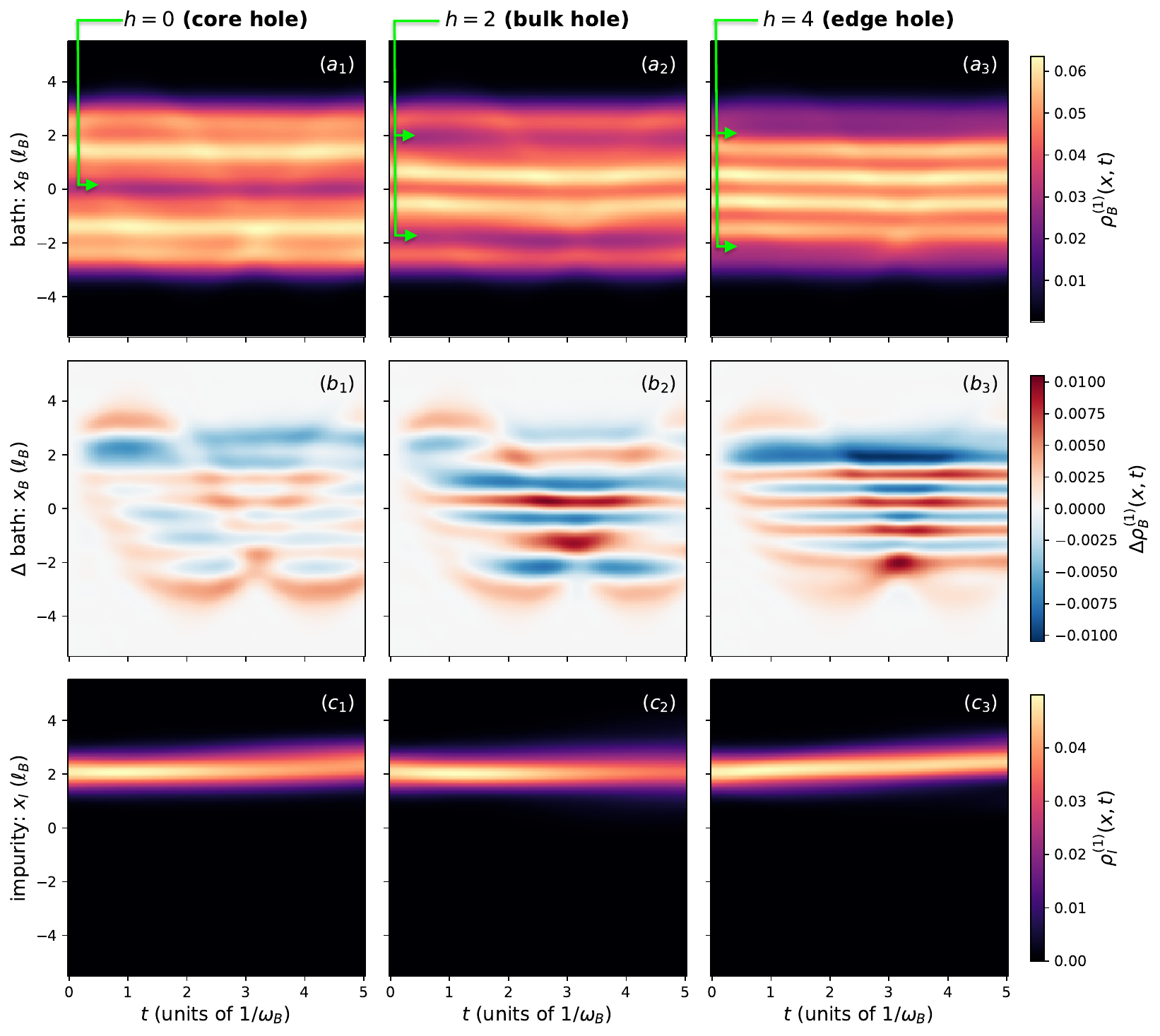}};

  \begin{scope}[x={(img.south east)}, y={(img.north west)}]

    \draw[myline]  (0.0650,0.9815) -- (0.1001,0.9815);
    \draw[myarrow] (0.0650,0.9815) -- (0.0650,0.818) -- (0.0862,0.818);

    \draw[myline]  (0.3418,0.9801) -- (0.3809,0.9801);
    \draw[myline]  (0.3418,0.9801) -- (0.3418,0.7704);
    \draw[myarrow] (0.3418,0.8701) -- (0.3684,0.8701);
    \draw[myarrow] (0.3418,0.7724) -- (0.3684,0.7724);

    \draw[myline]  (0.6180,0.9801) -- (0.653,0.9801);
    \draw[myline]  (0.6180,0.9801) -- (0.6180,0.7665);
    \draw[myarrow] (0.6180,0.8730) -- (0.6506,0.8730);
    \draw[myarrow] (0.6180,0.7665) -- (0.6506,0.7665);


  \end{scope}
\end{tikzpicture}
    \caption{Spatiotemporal evolution of the coupled bath-impurity dynamics. The top row \((a_i)\) displays the bath one-body density \(\rho_B^{(1)}(x;t)\), the middle row \((b_i)\) the associated density difference \(\Delta\rho_B^{(1)}(x;t)\), and the bottom row \((c_i)\) the impurity one-body density \(\rho_I^{(1)}(x;t)\). Parameters are \(g=1.0\,\hbar \ell_B \omega_B\), \(x_s=2\ell_B\), \(m_I=5.0m_B\), and \(\omega_I=0.5\omega_B\). From left to right, the columns correspond to the initial hole states \(h=0\) (\(i=1\), core hole), \(h=2\) (\(i=2\), bulk hole), and \(h=4\) (\(i=3\), edge hole). The green arrows indicate the density-depleted regions associated with hole initialization.}
    \label{fig:density_g10_xs2}
\end{figure*}
\begin{figure*}[t]
    \centering
    \includegraphics[width=\textwidth]{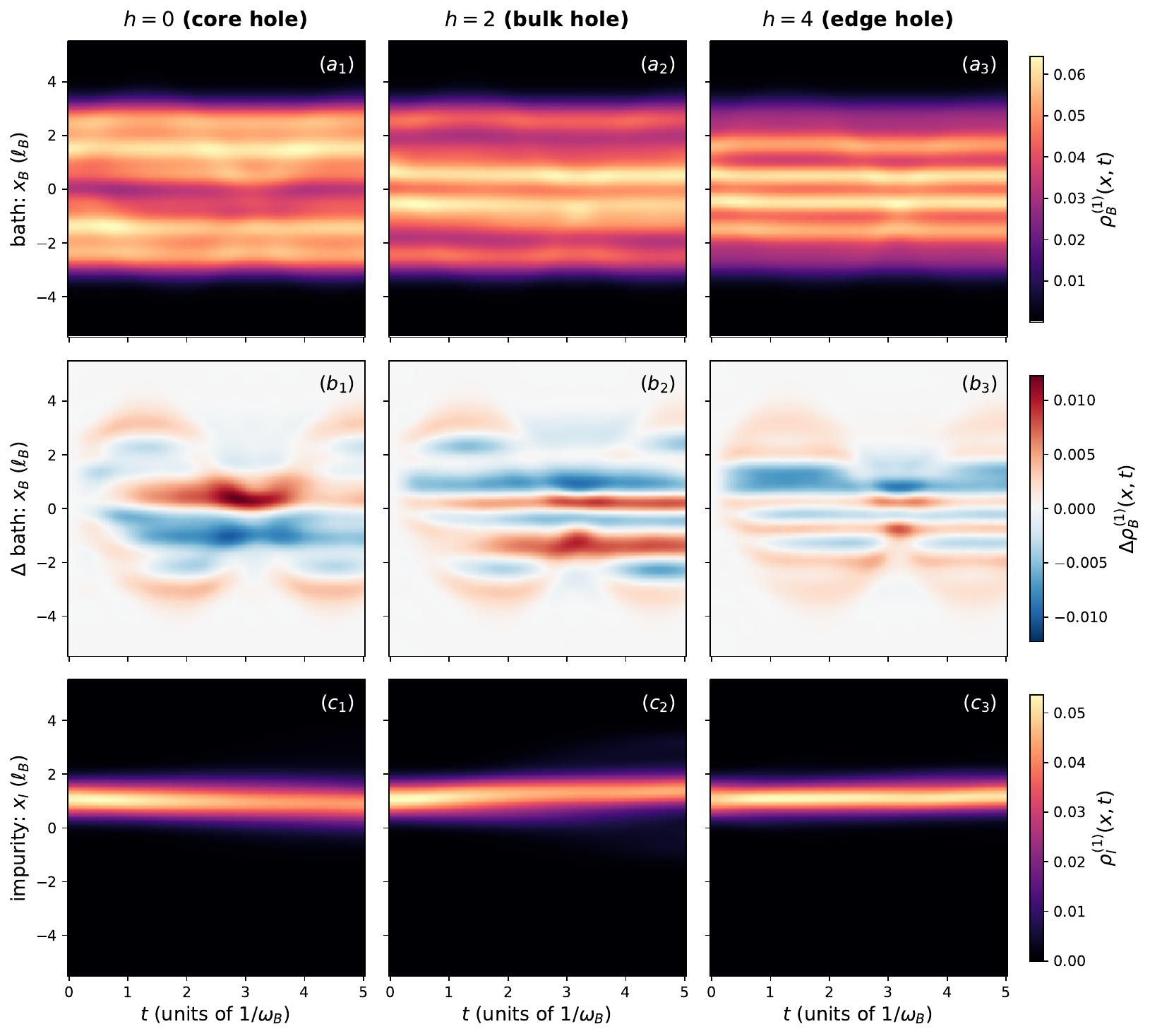}
    \caption{Spatiotemporal evolution of the coupled bath-impurity dynamics. The top row \((a_i)\) displays the bath one-body density \(\rho_B^{(1)}(x;t)\), the middle row \((b_i)\) the associated density difference \(\Delta\rho_B^{(1)}(x;t)\), and the bottom row \((c_i)\) the impurity one-body density \(\rho_I^{(1)}(x;t)\). Parameters are \(g=1.0\,\hbar \ell_B \omega_B\), \(x_s=1\ell_B\), \(m_I=5.0m_B\), and \(\omega_I=0.5\omega_B\). From left to right, the columns correspond to the initial hole states \(h=0\) (\(i=1\), core hole), \(h=2\) (\(i=2\), bulk hole), and \(h=4\) (\(i=3\), edge hole).}
    \label{fig:density_g10_xs1}
\end{figure*}
\begin{figure*}[t]
    \centering
    \includegraphics[width=\textwidth]{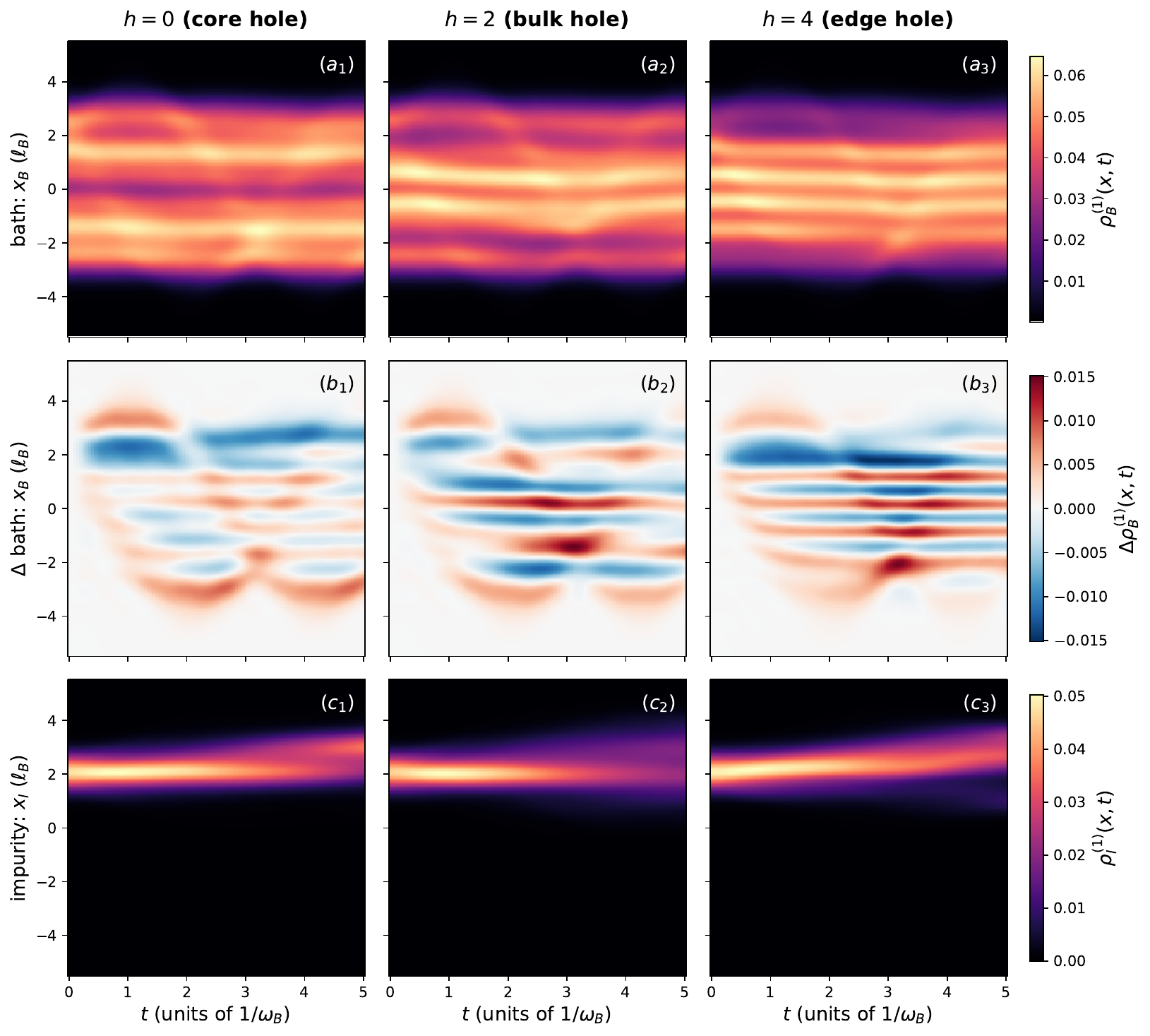}
    \caption{Spatiotemporal evolution of the coupled bath-impurity dynamics. The top row \((a_i)\) displays the bath one-body density \(\rho_B^{(1)}(x;t)\), the middle row \((b_i)\) the associated density difference \(\Delta\rho_B^{(1)}(x;t)\), and the bottom row \((c_i)\) the impurity one-body density \(\rho_I^{(1)}(x;t)\). Parameters are \(g=2.0\,\hbar \ell_B \omega_B\), \(x_s=2\ell_B\), \(m_I=5.0m_B\), and \(\omega_I=0.5\omega_B\). From left to right, the columns correspond to the initial hole states \(h=0\) (\(i=1\), core hole), \(h=2\) (\(i=2\), bulk hole), and \(h=4\) (\(i=3\), edge hole).}
    \label{fig:density_g20_xs2}
\end{figure*}
\begin{figure*}[t]
    \centering
    \includegraphics[width=\textwidth]{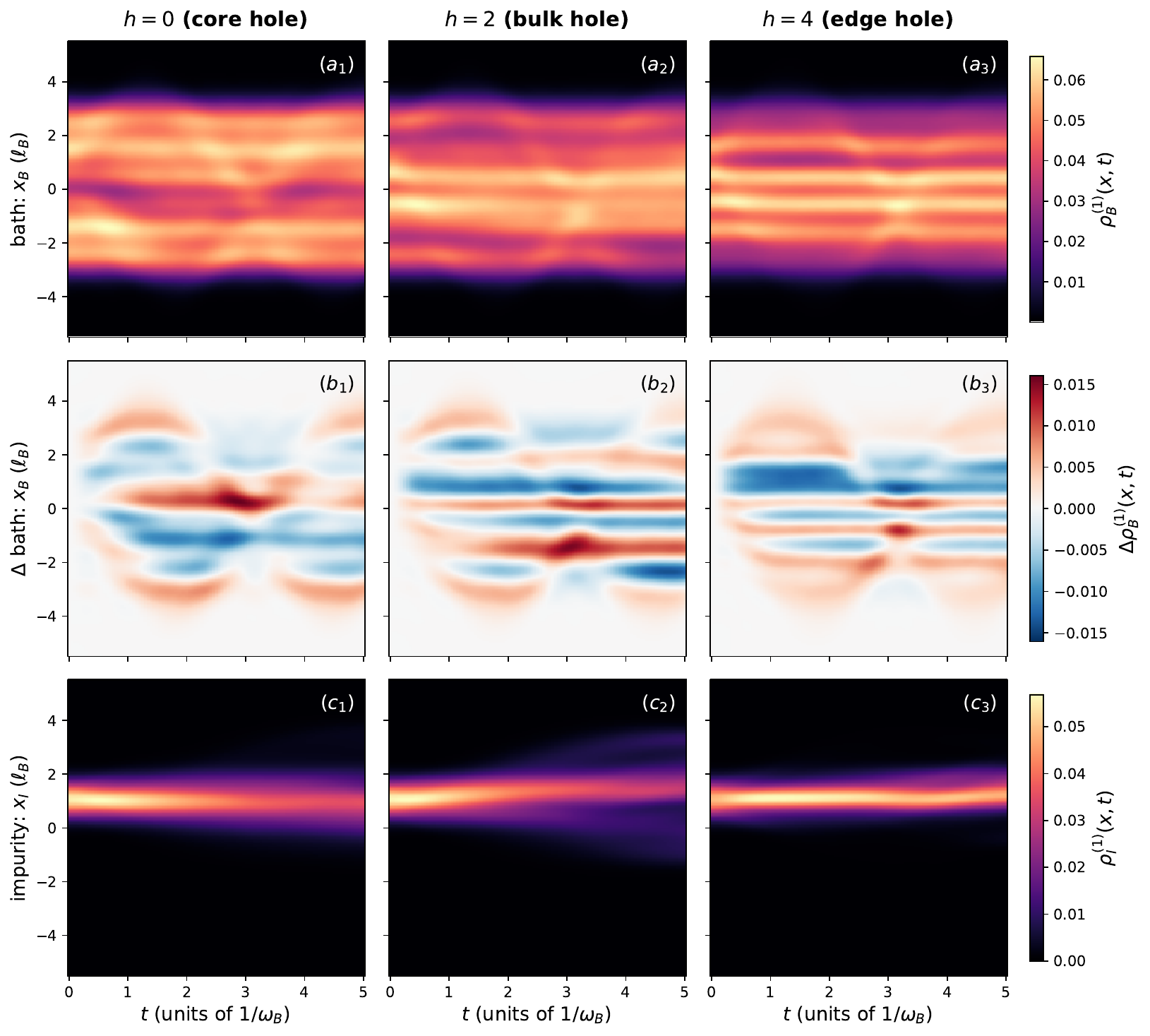}
    \caption{Spatiotemporal evolution of the coupled bath-impurity dynamics. The top row \((a_i)\) displays the bath one-body density \(\rho_B^{(1)}(x;t)\), the middle row \((b_i)\) the associated density difference \(\Delta\rho_B^{(1)}(x;t)\), and the bottom row \((c_i)\) the impurity one-body density \(\rho_I^{(1)}(x;t)\). Parameters are \(g=2.0\,\hbar \ell_B \omega_B\), \(x_s=1\ell_B\), \(m_I=5.0m_B\), and \(\omega_I=0.5\omega_B\). From left to right, the columns correspond to the initial hole states \(h=0\) (\(i=1\), core hole), \(h=2\) (\(i=2\), bulk hole), and \(h=4\) (\(i=3\), edge hole).}
    \label{fig:density_g20_xs1}
\end{figure*}
We begin by analyzing the spatiotemporal one-body densities of the bath and impurity species, which provide direct insight into how the impurity motion and the initially prepared bath hole shape the real-space dynamics. First, we consider the intermediate-interaction regime, $g=1.0\,\hbar\omega_B\ell_B$, with the impurity initially displaced to $x_s=2.0\ell_B$, i.e., positioned at the edge of the bath cloud as shown in Fig.~\ref{fig:density_g10_xs2}. This setup is designed to probe how the impurity perturbatively affects the bath, with the aim of inducing only weak excitations rather than a strong global response. For the core-hole case, $h=0$, switching on the interaction at $t=0$ slightly pushes the impurity further outward, signaling an initial demixing tendency between impurity and bath species. At the same time, the bath develops a weak density depletion propagating across the entire cloud, akin to a phonon mode, which becomes particularly visible in the density difference [panel (b$_1$)]. The impurity density also broadens while being displaced. A qualitatively similar behavior is observed for the bulk-hole preparation, $h=2$, where the bath oscillation pattern closely resembles that of the core-hole case, but appears to be more pronounced. In contrast, for the edge-hole case, $h=4$, the perturbation is weaker: although the bath still exhibits a similar oscillatory pattern, the impurity density remains more localized and shows no pronounced broadening. Overall, the main observation in this weakly perturbed regime is that the collective bath oscillation is largely insensitive to the specific hole preparation and exhibits a similar pattern in all cases, as can be inferred in particular from the bath density difference $\Delta\rho^{(1)}_B(x;t)$, see Fig.~\ref{fig:density_g10_xs2}~(b$_3$).

Next, we reduce the initial displacement to $x_s=1.0\ell_B$ while keeping the interaction strength fixed, as shown in Fig.~\ref{fig:density_g10_xs1}. In this case, the impurity is initially closer to the region of maximal bath density and therefore a stronger local response occurs. Indeed, the bath displays a more pronounced collective oscillation. For the core-hole case, the density dynamics clearly show a temporary refilling of the central depleted region during the oscillation, reflecting the stronger overlap of the impurity with the hole location. By contrast, for bulk- and edge-hole preparations, $h=2$ and $h=4$, the dynamical changes in bath and impurity densities become less pronounced. This can be traced back to the different spatial profiles of the involved bath orbitals depending on the hole position: the $h=0$ orbital has its maximum density at the trap center, whereas the higher orbitals corresponding to $h=2$ and $h=4$ are more extended and exhibit substantial probability density away from the center. Therefore, an impurity initially placed at small displacement overlaps most selectively with the core-hole configuration, while its effect on the bulk- and edge-hole preparations is spatially less distinct.

We now turn to the stronger interaction regime, $g=2.0\,\hbar\omega_B\ell_B$. For $x_s=2.0\ell_B$, shown in Fig.~\ref{fig:density_g20_xs2}, the excitation pattern in the bath becomes significantly more pronounced for all hole preparations. In the case of the core hole, $h=0$, the impurity is displaced more strongly and develops a visible internal minimum in its density profile. A similar feature also appears for the bulk-hole case $h=2$. More generally, a stronger bath-impurity coupling leads to more pronounced oscillatory dynamics of the bath, consistent with the previously identified phonon-like density-wave response in Fig.~\ref{fig:density_g10_xs2}~(a$_1$) and (b$_1$) for $h=0$. In particular, for $h=2$, the impurity density exhibits particularly strong broadening, indicating that this combination of interaction strength and initial displacement is especially sensitive to the spatial structure of the bath orbital.

Finally, we consider the stronger coupling together with the smaller displacement, $x_s=1.0\ell_B$, as shown in Fig.~\ref{fig:density_g20_xs1}. In this regime, all hole preparations exhibit pronounced collective bath excitations. For core-hole preparation, the impurity is drawn to $x_I = 0$, i.e., toward the location of the hole. For the other two hole preparation schemes in Fig.~\ref{fig:density_g20_xs1}~(c$_2$) and (c$_3$), in contrast to the weaker-coupling cases, the previously observed demixing tendency is strongly reduced: for the bulk-hole case, $h=2$, a slight residual demixing remains visible, but the most prominent feature is again the substantial broadening of the impurity density. This behavior is likely related to the density minima associated with the bath-hole structure. By contrast, for the edge-hole case, $h=4$, the impurity density remains comparatively compact and does not show a similar strong broadening. This is consistent with the flatter spatial structure of the edge orbital and the corresponding bath density distribution.

Having established these general features from the density evolution, we next turn to the mean position of the impurity to quantify the observed mixing and demixing behavior and to characterize the impurity motion more systematically.
\subsection{Impurity motion: mixing and demixing dynamics}
\label{subsec:results_ximean}
\begin{figure*}[t]
    \centering
    \includegraphics[width=\textwidth]{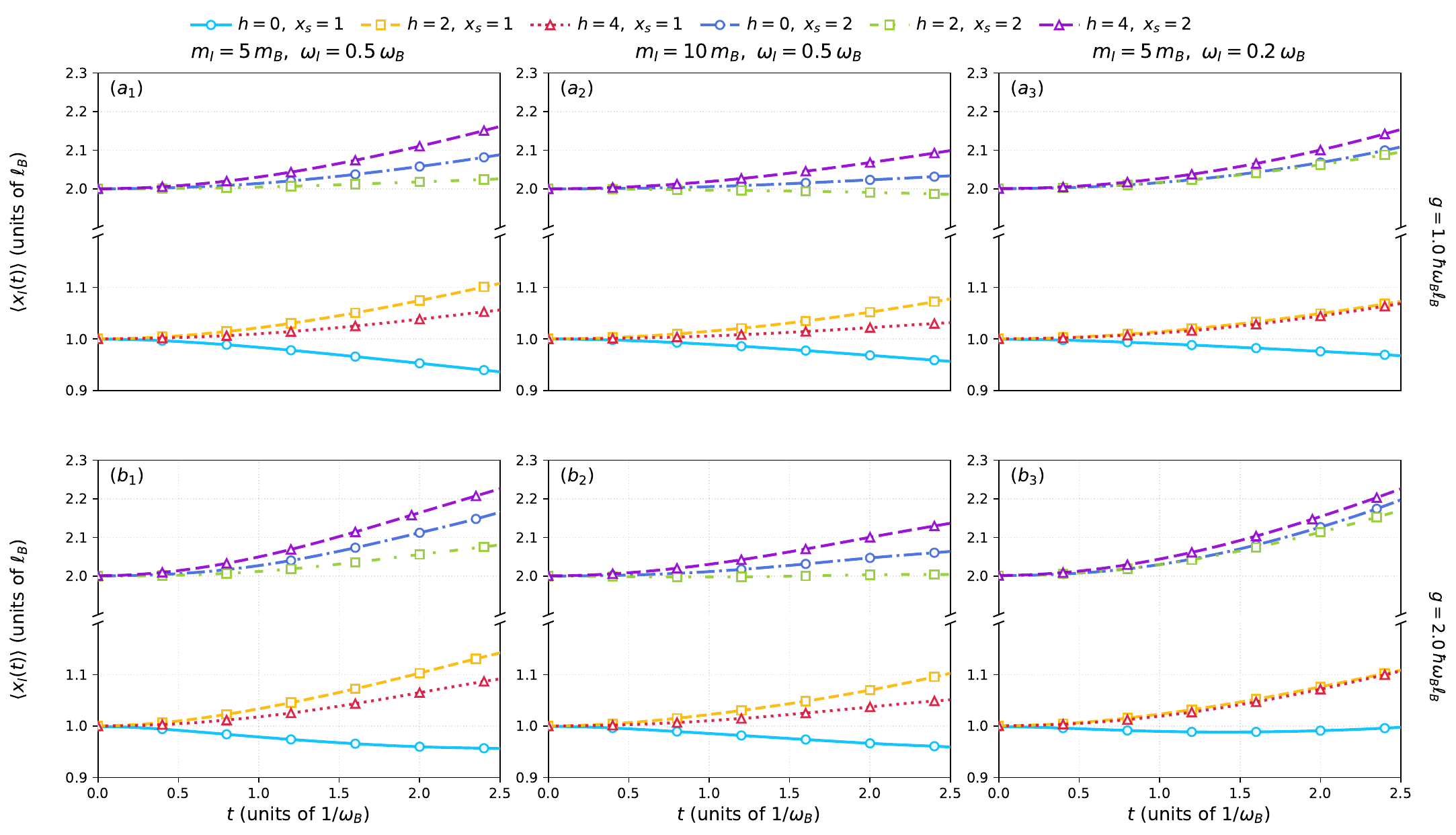}
    \caption{Impurity center-of-mass motion \(\langle x_I(t)\rangle\) for the three hole initial states \(h=0,2,4\). The upper (a$_i$) and lower (b$_i$) rows correspond to \(g=1.0\hbar\omega_B\ell_B\) and \(g=2.0\hbar\omega_B\ell_B\), respectively. The broken-axis representation enables a comparison of the trajectories for \(x_s=1\ell_B\) and \(x_s=2\ell_B\). The index \(i=1\) labels the reference setup with \(m_I=5.0m_B\) and \(\omega_I=0.5\omega_B\), while \(i=2\) and \(i=3\) correspond to the heavier-impurity case \(m_I=10m_B\) and the weaker-trap case \(\omega_I=0.2\omega_B\), respectively.}
    \label{fig:ximean}
\end{figure*}
To quantify the impurity dynamics beyond the qualitative picture provided by the density evolution, we now analyze the impurity center-of-mass position, $\langle x_I(t)\rangle$, shown in Fig.~\ref{fig:ximean}. This observable provides a direct measure of the degree of mixing or demixing between impurity and bath during the postquench dynamics. Across the individual panels, we compare the effect of the initial impurity displacement and the hole preparation, while the columns further illustrate how these trends are modified by changing the impurity mass and trapping frequency. In particular, moving row-wise at fixed interaction strength allows us to examine how reduced mobility caused by a larger mass, or enhanced mobility caused by weaker confinement, modifies the impurity response.

We first focus on the weak-coupling regime, $g=1.0\,\hbar\omega_B\ell_B$. For the smaller initial displacement, $x_s=1\ell_B$, shown in the left panels of Fig.~\ref{fig:ximean}, the core-hole preparation, $h=0$, leads to a clear inward motion of the impurity, indicating mixing with the bath. By contrast, for the bulk- and edge-hole cases, $h=2$ and $h=4$, the impurity is displaced further outward, signaling demixing. Among these, the strongest demixing occurs for $h=2$, which can be traced back to the density depletion associated with the hole in the bath profile. For the larger displacement, $x_s=2\ell_B$, the behavior changes qualitatively. In the core-hole case, the impurity no longer moves inward, consistent with the fact that the central hole cannot be efficiently refilled when the impurity starts farther away. At the same time, the demixing in the bulk-hole case becomes weaker, since the impurity is initially positioned closer to the corresponding depleted region. For the edge-hole preparation, $h=4$, the impurity is pushed farthest outward, reflecting the extended spatial character of the edge orbital.

These trends are modified quantitatively but not qualitatively, when increasing the impurity mass to $m_I=10m_B$. As expected, the heavier impurity exhibits a reduced overall displacement, consistent with its lower mobility. For the smaller shift, the same ordering between the different hole initial states remains visible, although the inward motion for $h=0$ is weakened. More strikingly, the larger shift and $h=2$ initial state supports a tendency toward mixing rather than demixing, indicating that the interplay between reduced mobility and the spatial structure of the hole can qualitatively alter the impurity trajectory. A similarly enhanced sensitivity is observed when weakening the impurity confinement. In this case, the impurity becomes more mobile and correspondingly more responsive to the detailed bath structure.

We now turn to the stronger interaction strength, $g=2.0\,\hbar\omega_B\ell_B$, shown in the lower panels, Fig.~\ref{fig:ximean}(b$_i$). Overall, the same qualitative features of the dynamics persist, but the balance between mixing and demixing shifts toward a stronger outward repulsion. For the reference setup, $m_I=5.0m_B$ and $\omega_I=0.5\omega_B$, the stronger coupling largely acts as a renormalization of the weak-coupling behavior: the core-hole case still shows the strongest tendency toward mixing, but the inward motion is reduced, while the demixing in the other hole initial states becomes more pronounced. Physically, the stronger repulsive interaction keeps the impurity more localized and suppresses its penetration into the bath.

A similar pattern is found for the heavier impurity, $m_I=10m_B$, although hereby the dynamics becomes even more restricted. In particular, for the bulk-hole case the partial mixing tendency observed at weaker coupling is suppressed, and the impurity remains more strongly localized near its initial position. Thus, enhanced interaction and reduced mobility cooperate to inhibit inward motion.

Finally, for the weak-trap case, Fig.~\ref{fig:ximean}(b$_3$), the stronger interaction favors demixing even more clearly. Most notably, the mixing tendency of the core-hole preparation at small initial displacement is suppressed. This shows that reduced spatial localization of the impurity, caused by the weaker confinement, facilitates a broader overlap with the bath and thereby lowers the energetic cost of an outward displacement. As a result, in all noncentral hole preparations the impurity is more efficiently expelled from the bath region.

In general, the center-of-mass motion of the impurity confirms and sharpens the picture suggested by the densities: mixing is favored primarily for the preparation of the core-hole and the small initial displacement, while demixing dominates for larger shifts, stronger interactions, and parameter regimes that enhance the susceptibility of the impurity to repulsive bath-induced forces.

After analyzing the mixing and demixing on the one-body level, we focus on the many-body correlations generated during the quench. To address this aspect, we investigate the impurity-bath entanglement through the von Neumann entropy.

\subsection{Entanglement growth and the special role of core-hole excitations}
\label{subsec:results_svn}

\begin{figure*}[t]
    \centering
    \includegraphics[width=\textwidth]{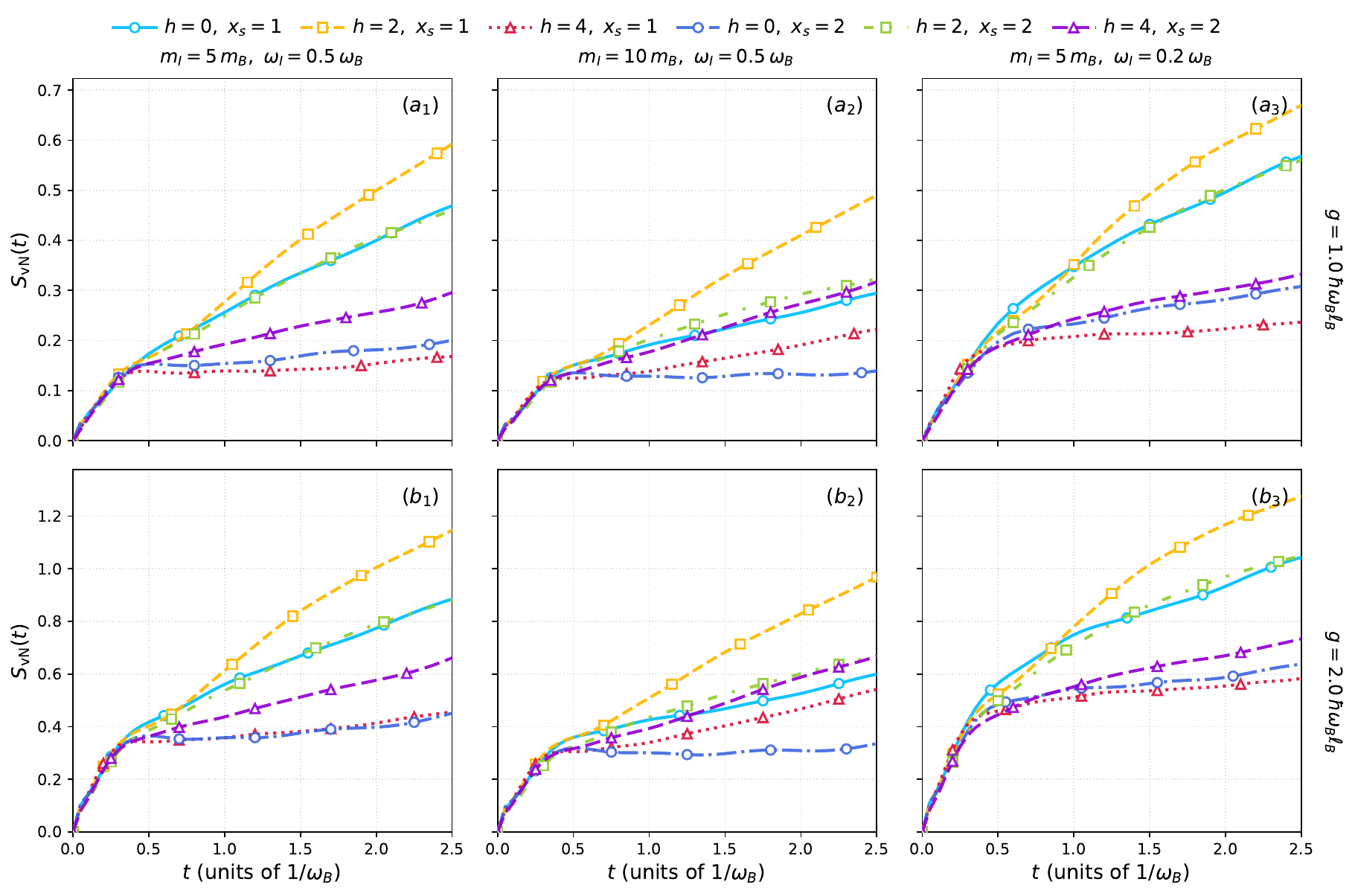}
    \caption{Time evolution of the impurity-bath von Neumann entropy \(S_{\mathrm{vN}}(t)\) for the three hole initial states \(h=0,2,4\). The upper (a$_i$) and lower (b$_i$) rows correspond to \(g=1.0\hbar\omega_B\ell_B\) and \(g=2.0\hbar\omega_B\ell_B\), respectively. Both values of shift \(x_s=1\ell_B\) and \(x_s=2\ell_B\) are provided. The index \(i=1\) labels the reference setup with \(m_I=5.0m_B\) and \(\omega_I=0.5\omega_B\), while \(i=2\) and \(i=3\) correspond to the heavier-impurity case \(m_I=10m_B\) and the weaker-trap case \(\omega_I=0.2\omega_B\), respectively.}
    \label{fig:svn}
\end{figure*}

As shown in Fig.~\ref{fig:svn}, the von Neumann entropy exhibits an overall initial increase over the first time interval, $\Delta t \approx 0.4/\omega_B$, for all investigated configurations. This early-time behavior is largely governed by the interaction quench itself and is consequently common to all setups. Comparing the upper panels (a$_i$), corresponding to the weaker interaction strength $g=1.0\hbar\omega_B\ell_B$, we find that the duration and magnitude of this initial collective growth depend on the impurity mass $m_I$ and trapping frequency $\omega_I$. In particular, the reference setup and the heavier impurity case, \(m_I=10\,m_B\), show an almost identical initial rise and saturation of the von Neumann entropy. Thus, increasing the impurity mass does not noticeably modify the early-time entanglement dynamics. However, a different behavior is observed for the more weakly trapped impurity (\(\omega_I=0.2\omega_B\)) [panel (a$_3$)], where the initial increase is more pronounced, whereas the splitting of the entropy curves sets in at nearly the same time. This behavior indicates that weaker confinement promotes a larger initial buildup of entanglement and extends the common early-time response before the details of the hole preparation become relevant.

After this initial collective stage, the entropy dynamics becomes strongly dependent on the interplay between hole position and initial impurity shift. For the reference case $m_I=5.0\,m_B$ and $\omega_I=0.5\omega_B$ [panel (a$_1$)], this dependence is particularly pronounced. For the smaller shift, $x_s=1.0\ell_B$, the Fermi-edge hole ($h=4$) leads to a strong suppression of the subsequent entanglement growth, such that the von Neumann entropy remains nearly constant at later times. In contrast, the core ($h=0$) and bulk ($h=2$) holes show a continued increase. This suggests that the corresponding orbitals provide a larger effective interaction region with the impurity. For the larger shift, $x_s=2.0\ell_B$, the trend changes: here the core-hole configuration ($h=0$) displays the weakest further increase of the entropy. This can be attributed to the large spatial separation between the deeply occupied core orbital and the initially displaced impurity, which reduces their effective overlap. In this case, the impurity is pushed further outward and couples only weakly to the bath background. By contrast, the bulk-hole case ($h=2$) still supports a broader interaction region and consequently yields the strongest entanglement growth. Indeed, for both shifts the largest entropy values are reached in the bulk-hole configuration.

Increasing the impurity mass to $m_I=10\,m_B$ [panel (a$_2$)] generally reduces the hole-dependent differences in the entropy dynamics. Accordingly, the separation between the individual curves becomes smaller. Most configurations show a reduced entanglement growth compared to the lighter-impurity case, consistent with the slower and more localized impurity motion. An exception occurs for the edge-hole case $h=4$ at $x_s=1\ell_B$, where the entropy increases more strongly. Similarly, for the larger shift $x_s=2\ell_B$, the edge-hole behavior remains comparable to that of the lighter impurity. For the bulk hole, $h=2$, the large-shift case remains close to the results of the reference setup, whereas the smaller-shift case exhibits a slight increase. The strongest reduction occurs for the core hole, $h=0$, where the entropy is clearly suppressed. Overall, the larger impurity mass reduces the impurity mobility and thus weakens its coupling to the bath, especially for the more localized core orbital, while the broader higher-lying orbitals remain more effective in generating entanglement.

A qualitatively opposite tendency is found when the impurity confinement is weakened to $\omega_I=0.2\omega_B$ [panel (a$_3$)]. The weaker trap increases the spatial extent of the impurity wave packet and enhances its mobility, which in general leads to stronger entanglement. In particular, for the core-hole case \(h=0\), the weaker confinement gives rise to a broader and more mobile impurity cloud. The decisive effect in the present case, however, is its increased spatial extent, which maintains a substantial overlap with the hole-prepared bath configuration even for the larger shift \(x_s=2\ell_B\). In the more strongly confined cases, this overlap is strongly suppressed. Consequently, the corresponding von Neumann entropy shows a stronger growth. This shows that the buildup of entanglement is governed not only by the impurity dynamics, but also crucially by the spatial shape of its wave packet.

The above observations allow for a simple physical interpretation. Increasing the impurity mass primarily suppresses its motion and consequently reduces the buildup of interspecies entanglement. Conversely, enhanced impurity mobility tends to favor stronger entanglement growth. At the same time, the results clearly demonstrate that the spatial overlap between the impurity and the unoccupied hole state is the key ingredient controlling the efficiency of the postquench excitation processes.

Turning to the stronger interaction regime ($g=2\hbar\omega_B\ell_B$), shown in the lower panels (b$_i$), we observe an overall increase of the entanglement compared to the weaker-coupling case. At the same time, the qualitative trends established for the intermediate interaction  $g=1\hbar\omega_B\ell_B$ with respect to impurity mass and trapping frequency remain intact. In particular, for the larger shift $x_s=2\ell_B$, the core-hole configuration $h=0$ consistently exhibits a comparatively reduced entropy growth across all parameter sets. This again reflects the spatial separation between the impurity and the corresponding hole state: the bath density effectively screens the impurity from the core-hole excitation, thereby suppressing the corresponding entanglement generation relative to the other setups.

Although the von Neumann entropy provides a global measure of the correlation buildup, it does not directly reveal the fate of the initially prepared vacancy. To address this central question, we now turn to the occupation $n_h(t)$ of the emptied bath orbital, which offers the most direct probe of hole persistence and refilling.
\subsection{Hole occupation and the emergence of core-hole excitations}
\label{subsec:results_hole}
\begin{figure*}[t]
    \centering
    \includegraphics[width=\textwidth]{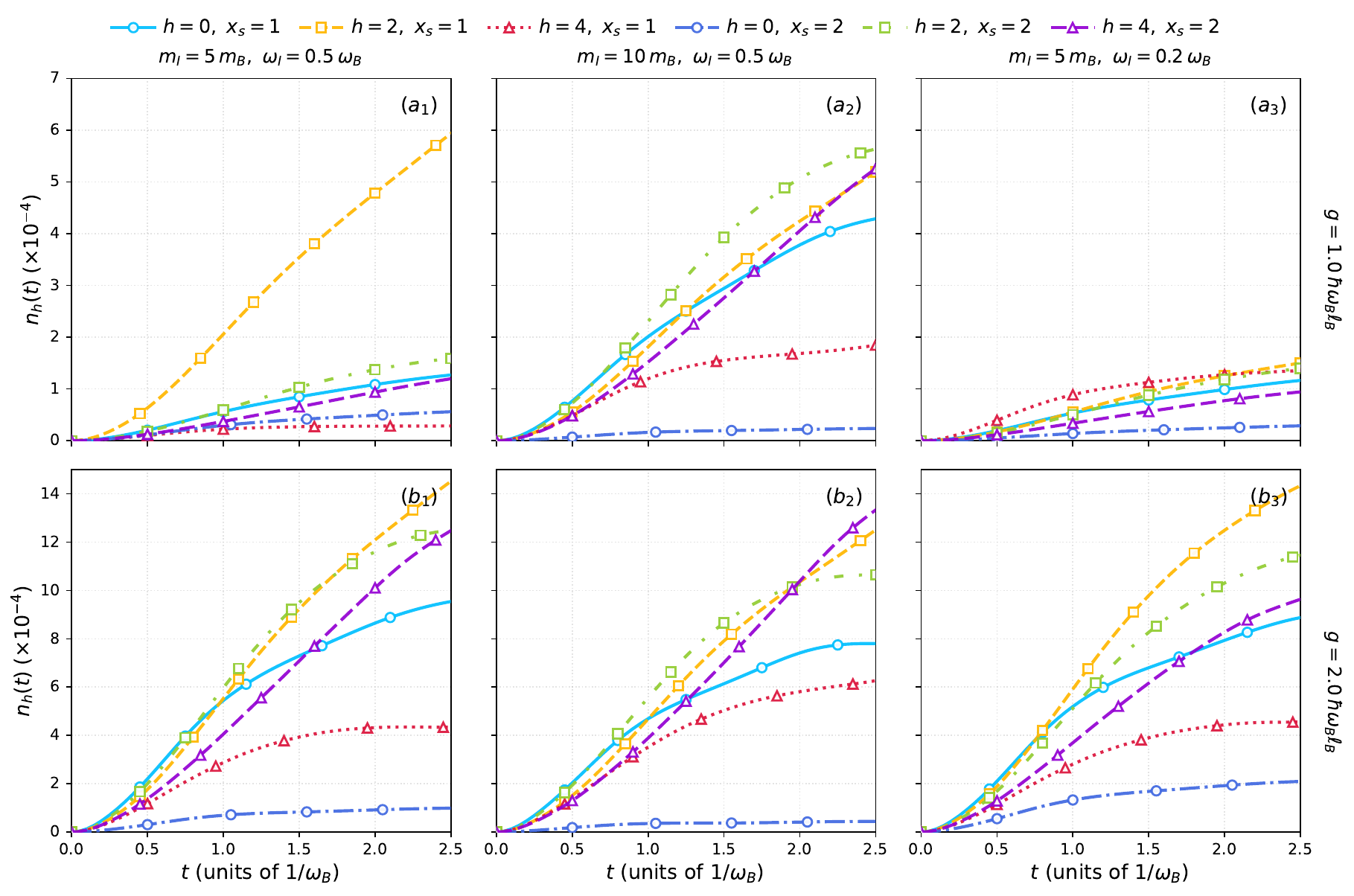}
    \caption{Time evolution of the hole-state occupation \(n_h(t)\) for the three hole initial states \(h=0,2,4\). The upper (a$_i$) and lower (b$_i$) rows correspond to \(g=1.0\hbar\omega_B\ell_B\) and \(g=2.0\hbar\omega_B\ell_B\), respectively. Both shift values \(x_s=1\ell_B\) and \(x_s=2\ell_B\) are provided. The index \(i=1\) labels the reference setup with \(m_I=5.0m_B\) and \(\omega_I=0.5\omega_B\), while \(i=2\) and \(i=3\) correspond to the heavier-impurity case \(m_I=10m_B\) and the weaker-trap case \(\omega_I=0.2\omega_B\), respectively.}
    \label{fig:hole_occ}
\end{figure*}
Considering the occupation of the hole state, $n_h(t)$ [Eq.\eqref{eq:nh_def}], this observable provides the most direct summary of the postquench many-body response, since it condenses the effects of impurity-induced density redistribution, orthogonality dynamics, and correlation buildup into the fate of the initially emptied bath orbital. Therefore, we investigate the behavior of hole occupation in Fig.~\ref{fig:hole_occ}. 

The upper panels of Fig.~\ref{fig:hole_occ}(a$_i$) show the weaker interaction regime, $g=1.0\hbar\omega_B\ell_B$, while the lower panels (b$_i$) corresponds to the stronger coupling, $g=2.0\hbar\omega_B\ell_B$. The first central observation is that increasing the interaction strength systematically enhances the refilling of the hole orbital, demonstrating that the interaction is the dominant driving mechanism behind the hole-state occupation. This trend is fully consistent with the stronger orthogonality response and the enhanced entanglement generation discussed previously in Fig.~\ref{fig:svn}: stronger coupling promotes a more substantial rearrangement of the bath around the impurity and, thereby, a more efficient redistribution of population into the initially vacant mode. However, at the level of the impurity parameters, the magnitude of $n_h(t)$ is strongly modulated by the impurity mass $m_I$ and in particular the trapping frequency $\omega_I$, which determine the spatial overlap between impurity and bath and thus, the efficiency of the induced transfer process. In particular, the combination of a core hole, $h=0$, with the larger displacement, $x_s=2.0\ell_B$, yields the smallest occupation overall, indicating that strong geometric separation suppresses the refill of the emptied orbital even when the orbital lies in the central, otherwise most relevant region of the bath. A similarly reduced occupation is found for the edge hole, $h=4$, with a small displacement, again reflecting that the refill process is highly sensitive to the detailed spatial arrangement of the impurity-hole rather than to the hole index alone. This behavior mirrors the structure already inferred from the entanglement $S_{vN}$ in Fig.~\ref{fig:svn}, where the largest increase in entanglement was found for those configurations that also favor enhanced coupling between the impurity and the hole-depleted bath region. Among impurity configurations, the reference and heavy-impurity cases [panels (a$_1$) and (a$_2$)] display broadly comparable dynamics, whereas the weak-confinement case [panel (a$_3$)] shows a markedly suppressed hole occupation in the weak-coupling regime. This identifies tight enough confinement, and hence a sufficiently localized impurity density near the bath, as an essential ingredient for substantial repopulation of the empty orbital. However, considering the stronger coupling $g=2.0\hbar\omega_B\ell_B$ the comparison between weak and strong interaction [especially Fig.~\ref{fig:hole_occ}(a$_3$) and (b$_3$)] shows that interaction effects can partially overcome this geometric disadvantage, such that even the weakly trapped impurity induces a noticeably larger occupation. 

Hence, $n_h(t)$ unifies the main conclusions of the preceding analysis: while the interaction strength sets the overall scale of the many-body response, the impurity mass and confinement geometry decisively shape how efficiently this response is converted into an actual refill of the hole state.
\section{Summary and Outlook}
\label{sec:conlcusion_and_outlook}

Summarizing, our results establish a general physical picture of hole dynamics in the interacting impurity-bath system. The density evolution and in particular the center-of-mass motion of the impurity demonstrate that the impurity acts as a mobile scatterer whose coupling to the fermionic bath is governed by the interplay of interaction strength, confinement, and the character of the hole, i.e., the initially prepared bath excitation, leading to a dynamical competition between mixing and demixing of the bath and impurity species. Going beyond this single-particle picture, the von Neumann entropy quantifies the entanglement generated during the evolution and shows that these processes are accompanied by pronounced many-body correlations. Finally, this collective response is manifested in the hole occupation, which directly tracks the fate of the initially emptied bath orbital and, most importantly, distinguishes deep core holes from bulk or edge vacancies by demonstrating their substantially enhanced dynamical stability. Hence, core-hole excitations emerge as a robust dynamical many-body feature, retaining their stability in a regime where impurity motion, bath rearrangement, entanglement growth, and orthogonality response are all clearly developed.

Future perspectives arise naturally from the present dynamical core-hole picture. An immediate extension is to address multiple impurities, spinful baths, or attractive interspecies interactions, where pairing and bound-state formation may compete with the refilling dynamics identified here. A further important direction is the crossover from the few-body trapped regime to larger fermionic systems and finite temperatures, which would establish a more direct link to the mesoscopic and thermodynamic formulations of the x-ray-edge problem. Because the essential ingredients of our protocol, which are given by mass imbalance, interaction quenches, and precise preparation of the initial state, are accessible in ultracold-atom experiments, our results also motivate time-resolved measurements of hole refilling and impurity motion as probes of nonequilibrium many-body dressing. In this way, core-hole excitations in ultracold gases may offer a new route toward engineering and observing robust correlated excitations beyond the conventional quasiparticle paradigm.
\section*{Acknowledgements}
This work has been funded by the Cluster of Excellence “Advanced Imaging of Matter” of the Deutsche
Forschungsgemeinschaft (DFG) - EXC 2056 - project ID 390715994. G.K.M. has received funding by the Austrian Science Fund (FWF) [DOI: 10.55776/F1004]. For this work the HPC-cluster Hummel-2 at University of Hamburg was used. The cluster was funded by Deutsche Forschungsgemeinschaft (DFG, German Research Foundation) – 498394658.
\appendix
\section{The ML-X method}
\label{app:MLX}
At the beginning of Sect.~\ref{sec:theory}, we provided a concise overview of the numerical approaches employed. The purpose of this section is to dive into the ML-X method.
The ML-X ansatz expresses the total many-body wavefunction, $|\Psi(t)\rangle$, as a linear combination of $j=1,2,\dots,D$ distinct orthonormal functions for each involved species
\begin{equation}
    |\Psi(t)\rangle = \sum_{j_B, j_I = 1}^D A_{j_B, j_I}(t) |\Psi_{j_B}^B(t)\rangle |\Psi_{j_I}^I(t)\rangle,
    \label{eqn:wavefunction_mlx}
\end{equation}
where $|\Psi_j^\sigma(t)\rangle$ ($\sigma = B, I$) are species wavefunctions, and $A_{j_B, j_I}(t)$ are time-dependent expansion coefficients. This decomposition is formally equivalent to a Schmidt decomposition of rank $D$
\begin{equation}
    |\Psi(t)\rangle = \sum_{k=1}^D \sqrt{\lambda_k(t)} |\tilde{\Psi}_k^B(t)\rangle |\tilde{\Psi}_k^I(t)\rangle,
\end{equation}
considering the $\lambda_k(t)$ (Schmidt weights) as eigenvalues of the reduced density matrix $\rho_\sigma^{(N_\sigma)}(t)$ and eigenstates $|\tilde{\Psi}_k^\sigma(t)\rangle$ as Schmidt modes.
Here, the reduced density matrix for species $\sigma$ is given by
\begin{equation}
\begin{split}
\rho_{\sigma}^{(N_\sigma)}&(x_1, \dots, x_{N_\sigma}, x'_1, \dots, x'_{N_\sigma}, t)=
\int \prod_{j = 1}^{N_{\bar \sigma}}\mathrm{d}  x_j^{\bar \sigma}~ \\
&\times \Psi^*(x_1^{\sigma}=x'_1, \dots, x_{N_\sigma}^{\sigma}=x'_{N_\sigma}, x^{\bar \sigma}_1, \dots, x^{\bar \sigma}_{N_{\bar \sigma}}, t) \\
&\times \Psi(x_1^{\sigma}=x_1, \dots, x_{N_\sigma}^{\sigma}=x_{N_\sigma}, x^{\bar \sigma}_1, \dots, x^{\bar \sigma}_{N_{\bar \sigma}}, t),
\end{split}
\end{equation}
with $\bar{\sigma} \neq \sigma$. Here, $N_{\sigma}$ and $N_{\bar{\sigma}}$ denote the number of atoms belonging to the respective species $\sigma$, $\bar{\sigma}$. Within the ML-X framework, the density matrix operator can be expanded as $\rho_{\sigma}^{(N_\sigma)}(x_1, \dots, x_{N_\sigma}, x'_1, \dots, x'_{N_\sigma}, t)=\langle x_1, \dots, x_{N_\sigma}  | \hat{\rho}^{(N_{\sigma})}_{\sigma}(t)| x'_1, \dots, x'_{N_\sigma} \rangle$ with the density matrix operator 
\begin{equation}
\hat{\rho}_\sigma^{(N_\sigma)}(t) = \sum_{j_\sigma,j'_\sigma=1 \atop j_{\bar{\sigma}}=1}^D \left[\hat{\rho}_\sigma^{(N_\sigma)}(t)\right]_{j_\sigma,j'_\sigma} |\Psi_{j_\sigma}^\sigma(t)\rangle \langle\Psi_{j'_\sigma}^\sigma(t)|
\end{equation}
with the matrix elements $\sum_{j_{\bar{\sigma}}} A_{j_{\sigma}, j_{\bar{\sigma}}}^*(t) A_{j'_{\sigma}, j_{\bar{\sigma}}}(t)\equiv \left[\hat{\rho}_{\sigma}^{(N_\sigma)}(t)\right]_{j_{\sigma}, j'_{\sigma}}$ calculated from the time-dependent expansion coefficients, see Eq.~\eqref{eqn:wavefunction_mlx}.
Hence, diagonalizing $\left[\hat{\rho}_\sigma^{(N_\sigma)}(t)\right]_{j_\sigma,j'_\sigma}$ for $j_{\sigma},j_{\sigma}'=1,\cdots,D$ yields $\lambda_k(t)$ and $|\tilde{\Psi}_k^\sigma(t)\rangle$.\\\\
The key feature of the ML-X method is the multilayered structure. It arises from the expansion of each species wavefunction, $|\Psi_j^\sigma(t)\rangle$ in terms of time-dependent number states $|\vec{n}(t)\rangle^\sigma$ leading to
\begin{equation}
    |\Psi_j^\sigma(t)\rangle = \sum_{\vec{n}} B_{j,\vec{n}}^\sigma(t) |\vec{n}(t)\rangle^\sigma,
\end{equation}
where $B_{j,\vec{n}}^\sigma(t)$ corresponds to time-dependent expansion coefficients in the single-species number state basis, $|\vec{n}(t)\rangle^\sigma$. These number states are built in terms of $d^\sigma$ time-dependent variationally optimized Single-Particle Functions (SPFs) given by $\phi_l^\sigma(t)$, $l=1,2,..., d^\sigma$ with $\vec{n}=(n_1,...,n_{d^{\sigma}})$ corresponding to the number of atoms in each SPF. In the lowest layer, the SPFs are expanded in a time-independent DVR basis $\{|k\rangle\}$ and are defined as
\begin{equation}
    |\phi_j^\sigma(t)\rangle = \sum_{k=1}^{\mathcal{M}} C_{jk}^\sigma(t) |k\rangle.
\end{equation}
In this study, we used the points on the DVR grid $\mathcal{M} = 150$ of a harmonic oscillator.\\\\
To compute the ground state at the beginning of our analysis, imaginary-time propagation is performed using $\tau = -it$. This causes the energy of the state to decay proportionally to $e^{-(E(t)-E_0)t}$, converging to the ground state ($E_0$) as $t \to \infty$.\\\\
The ansatz’s Hilbert space truncation is characterized by the choice of the orbital configuration space, which is represented by $C = (D; d^B; d^I)$, where we choose the following set for our investigation \cite{Becker2024}:
\begin{itemize}
    \item $d^B = 18$: Bath orbitals to capture the intra-species bath correlations
    \item $d^I = 12$: 
    Are found to be enough for convergence of the impurity species
    \item $D = d^I = 12$: Incorporating all possible Schmidt modes of inter-species entanglement, for given $d^I$. 
\end{itemize}
This choice is based on an extensive convergence analysis for our ground state, which has also demonstrated its validity in the dynamics, as shown by examining the occupation numbers of the respective orbitals.
The equations of motion are derived using the Dirac-Frenkel variational principle \cite{Dirac1930annihilation, frenkel1934wave}
\begin{equation}
    \langle \delta \Psi(t)| i\hbar \frac{\partial}{\partial t} - H |\Psi(t)\rangle = 0.
\end{equation}
This leads to $D^2$ linear differential equations for $A_{j_B,j_I}(t)$, coupled to nonlinear integro-differential equations for $B_{j,\vec{n}}^\sigma(t)$ and $C_{j,k}^\sigma(t)$.
\section{Derivation of potential-energy-surface diagnostics in second quantization}
\label{app:pes_diagnostics_second_quantization}
In this appendix, we provide explicit expressions for the geometric renormalization term entering the MCBO equations of motion, together with the second-quantized derivation of the impurity density matrix and the projector-based diagnostics used to analyze potential-energy-surface occupations.

The geometric renormalization term appearing in Eq.~\eqref{eqn:schroedinger-time-dendent} reads
\begin{equation}
V_{kl}^{\rm ren}(x_I)=\frac{\hbar^2}{2m_I}\left\langle\frac{\partial \Psi_{k,B}}{\partial x_I}(x_I)\middle|1-\hat{\mathcal P}_M\middle|\frac{\partial \Psi_{l,B}}{\partial x_I}(x_I)\right\rangle,
\label{eqn:potential_renormalization}
\end{equation}
where
\begin{equation}
\hat{\mathcal P}_M
=
\sum_{j=1}^{M}
|\Psi_{j,B}(x_I)\rangle \langle \Psi_{j,B}(x_I)|
\end{equation}
is the projector onto the truncated \(M\)-channel bath subspace. Physically, the potential renormalization represents how the motion of the impurity species causes changes in the kinetic energy of the bath, as well as subtle adjustments in the bath-impurity interaction that are not captured by the adiabatic approximation solely.

The multi-channel Born-Oppenheimer ansatz of Eq.~\eqref{eqn:multi-channel_BornOppenheimer} written in the second quantization formalism reads
\begin{equation}
|\Psi (t)\rangle = \sum_{j=1}^{M} \int d x_I \, \Psi_{j,I}(x_I; t) \hat{\Psi}_I^\dagger(x_I) |0_I\rangle \otimes |\Psi_{j,B}(x_I)\rangle.
\label{eqn:mcbo_2nd_quantize}
\end{equation}
Here, \( \Psi_{j,I}(x_I; t) \) denotes the channel-resolved impurity wavefunction and is the only time-dependent variational quantity in the ansatz. The states \( |\Psi_{j,B}(x_I)\rangle \) describe the bath for a fixed impurity position \(x_I\). As discussed in Sec.~\ref{mcBO_main_text}, they are chosen as eigenstates of the bath Hamiltonian  $\hat{H}_B + \hat{H}_{BI}$ with $x_I$ treated as an external parameter. Finally, the operator \( \hat{\Psi}_I^\dagger(x_I) \) creates an impurity particle at \(x_I\) and together with the corresponding annihilation operator \( \hat{\Psi}_I(x_I) \) obeys the canonical fermionic anti-commutation relation $\{\hat{\Psi}_I(x_1), \hat{\Psi}_I^\dagger(x_2)\} = \delta(x_1 - x_2)$.

\subsection{Impurity one-body density matrix}
In this formalism, the one-body density matrix of the impurity is defined as
\begin{equation}
\rho_I^{(1)}(x_1,x_2; t) = \langle \Psi(t)| \hat{\Psi}_I^\dagger(x_1) \hat{\Psi}_I(x_2) |\Psi(t)\rangle,
\end{equation}
which after some straightforward fermionic anti-commutation algebra reads
\begin{equation}
\begin{split}
    \rho_I^{(1)}(x_1,x_2;t) = \sum_{j,k=1}^M& \Psi_{j,I}^*(x_1;t)\Psi_{k,I}(x_2;t) \\ &\times \langle \Psi_{j,B}(x_1) | \Psi_{k,B}(x_2) \rangle.
\end{split}
\end{equation}
This shows that the one-body density matrix contains contributions from overlaps between bath states \( \langle \Psi_{j,B}(x_1) | \Psi_{k,B}(x_2) \rangle \), weighted by the corresponding impurity amplitudes. Notice that the one-body density $\rho^{(1)}_I(x_I;t) = \rho^{(1)}(x_I, x_I;t)$ further simplifies to 
\begin{equation}
\begin{split}
    \rho_I^{(1)}(x_I;t) &= \sum_{j = 1}^M \underbrace{|\Psi_{j,I}(x_I;t)|^2}_{\equiv \rho_{I,j}^{(1)}(x_I;t)},
\end{split}
\label{one_body_density_components}
\end{equation}
where we have defined $\rho_{j, I}^{(1)}(x_I;t)$ as the channel-resolved density contribution. The overlap between bath states is simplified because $\langle \Psi_{j,B}(x_I) | \Psi_{k,B}(x_I) \rangle = \delta_{j,k}$. 
\subsection{Projectors onto Born-Oppenheimer channels}
To derive the probability of occupying a certain PEC curve $j$, we introduce the projector 
\begin{equation}
\hat{\mathcal{P}}_j = \int d x_I \, |\Psi_j(x_I)\rangle \langle \Psi_j(x_I)|,
\end{equation}
where $|\Psi_j(x_I)\rangle$ are defined as
\begin{equation}
|\Psi_j(x_I)\rangle = \hat{\Psi}_I^\dagger(x_I)|0_I\rangle \otimes |\Psi_{j,B}(x_I)\rangle.
\end{equation}
This projector resolves the many-body state onto the $j$-th Born-Oppenheimer channel, integrated over all impurity positions. It is straightforward to verify that these operators satisfy the projector relations $\hat{\mathcal{P}}_{j}\hat{\mathcal{P}}_{k} = \delta_{j,k} \hat{\mathcal{P}}_{j}$ and $\sum_{j = 1}^M \hat{\mathcal{P}}_{j} = \hat{I}_{M}$, where $\hat{I}_M$ denotes the identity in the truncated $M$-channel subspace. These properties can be straightforwardly proven by noting that $\langle\Psi_j(x_1)|\Psi_k(x_2)\rangle = \delta_{j,k} \delta(x_1 - x_2)$, stemming from fermionic anti-commutation relations and $\langle \Psi_{j,B}(x_I) | \Psi_{k,B}(x_I) \rangle = \delta_{j,k}$ (see discussion below Eq.~\eqref{one_body_density_components}).
In the next step, we calculate the expectation value of the projector
\begin{equation}
\begin{split}
    \langle \Psi (t) | \hat{\mathcal{P}}_j | \Psi (t) \rangle &= \int d x_I \, |\Psi_{j,I}(x_I;t)|^2 \\&= \int d x_I \, \rho^{(1)}_{I,j}(x_I;t),
\end{split}
\label{projector_in_PEC}
\end{equation}
where we have again used $\langle \Psi_{j,B}(x_I) | \Psi_{k,B}(x_I) \rangle = \delta_{j,k}$.
The expression above shows that the probability of occupying a particular state on the PEC depends solely on the norm of the impurity wavefunction in the $j$-th channel integrated over all space. In addition, Eq.~\eqref{projector_in_PEC} justifies {\it a posteriori} the definition of $\rho^{(1)}_{j, I}(x_I)$ as the contribution of the $j$-th PEC to the one-body density in Eq.~\eqref{one_body_density_components}, since the integral of this density function gives the occupation of the corresponding PEC.

More specifically, one may further resolve the $j$-th channel into eigenstates \(| \phi_{m, j} \rangle\) of the impurity motion on that PEC.
\begin{equation}
\hat{\mathcal{P}}_{m,j} = |\Psi_{m,j}\rangle \langle \Psi_{m,j}|.
\end{equation}
The corresponding state in this case is defined as 
\begin{equation}
|\Psi_{m,j}\rangle = \int d x_I \, \phi_{m,j}(x_I) \hat{\Psi}_I^\dagger(x_I) |0_I\rangle \otimes |\Psi_{j,B}(x_I)\rangle,
\end{equation}
where $\phi_{m,j}(x_I)$ is the wavefunction of the impurity state of which we wish to determine the occupation. The operator $\hat{\mathcal{P}}_{m,j}$ obeys all the proper projector properties, provided that the $| \phi_{m, j} \rangle$ states for varying $m$ form an orthonormal basis, the particular property $\hat{\mathcal{P}}_{m,j}^2 = \hat{\mathcal{P}}_{m,j}$ only requires that the state is normalized $\int {\rm d}x_I~|\phi_{m,j}(x_I)|^2 = 1$.
By incorporating the above assumptions the expectation value of the projector $\hat{\mathcal{P}}_{m,j}$ reads
\begin{equation}
\langle \Psi (t) | \hat{\mathcal{P}}_{m,j} | \Psi (t) \rangle = \left| \int d x_I \, \Psi_{I,j}^*(x_I;t) \phi_{m,j}(x_I) \right|^2.
\end{equation}
Note that $\langle \Psi_{j,B}(x_I) | \Psi_{k,B}(x_I) \rangle = \delta_{j,k}$ was used for this derivation (see discussion below Eq.~\eqref{one_body_density_components}).
Therefore, the occupation of a given state \( \phi_{m,j}(x_I) \) of the $j$-th PEC in the many-body state, is just the squared overlap of $j$-th PEC component of the MCBO wavefunction $\Psi_{j,I}^*(x_I;t)$ with the wavefunction of the target state $\phi_{m,j}(x_I)$. 

\section{Convergence Analysis}
\label{app:Convergence}

\begin{figure*}[t]
    \centering
    \includegraphics[width=\textwidth]{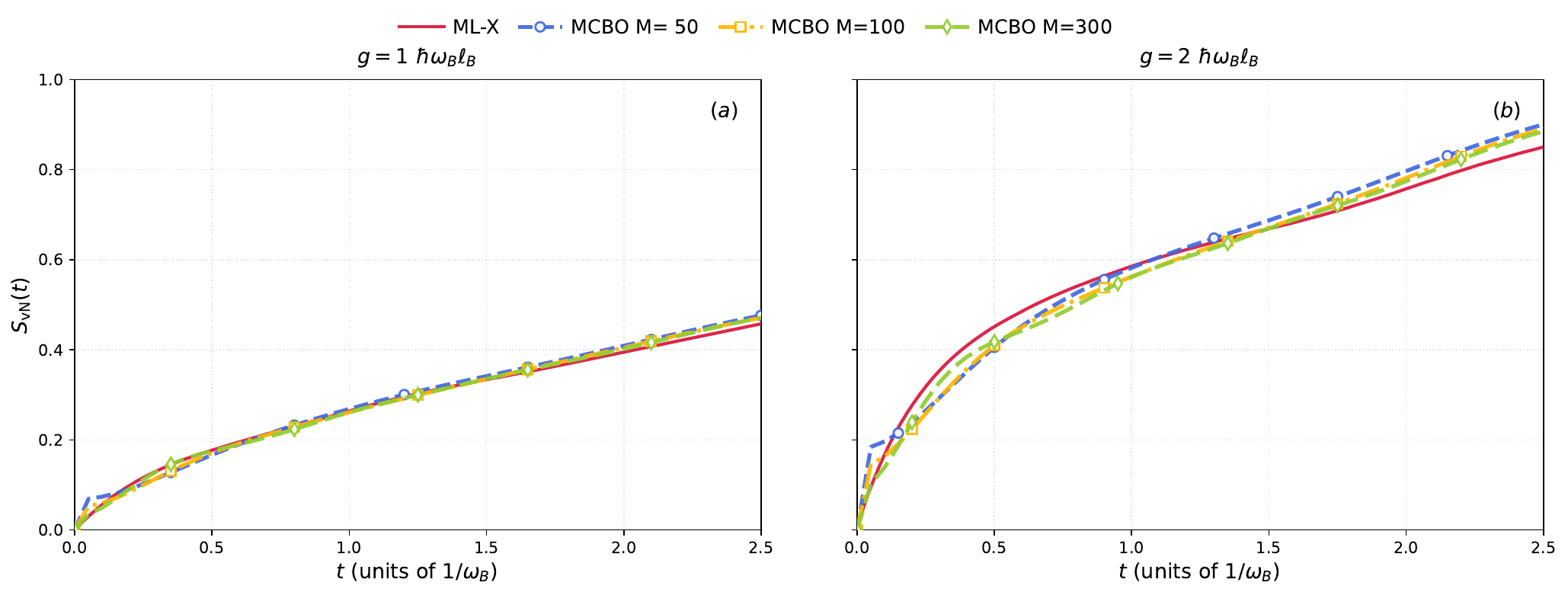}
    \caption{Time evolution of the impurity-bath von Neumann entropy \(S_{\mathrm{vN}}(t)\) for the core-hole initial state with \(h=0\). The left panel, (a), corresponds to the intermediate interaction strength \(g=1.0\,\hbar\omega_B\ell_B\), while the right panel, (b), shows the stronger interaction case \(g=2.0\,\hbar\omega_B\ell_B\). In each panel, the ML-X result is compared to MCBO calculations with different numbers of PECs.}
    \label{fig:sVN_convergence}
\end{figure*}

To verify convergence and justify the use of the MCBO method, we focus on the von Neumann entropy \(S_{\mathrm{vN}}\), which can be accessed straightforwardly within the ML-X approach. The convergence of the ML-X calculations has already been established in our previous works~\cite{Becker2024, Becker2025}. Since the present study addresses quench dynamics and places particular emphasis on entanglement, it is essential to demonstrate that the MCBO method also yields reliable results. In particular, as we have shown in Refs.~\cite{Becker2024, Becker2025} it is a sensitive measure of convergence of MCBO. In general, entanglement measures are more sensitive to convergence than one-body densities and related observables. Therefore, establishing convergence for the von Neumann entropy provides strong evidence for the reliability of the remaining quantities discussed in the main text.

To this end, Fig.~\ref{fig:sVN_convergence} compares MCBO results obtained with \(M=50\), \(100\), and \(300\) PECs to the ML-X reference. In the main text, we use \(M=300\), while the ML-X calculations are performed with the orbital configuration space \(C=(D=12,\, d^B=18,\, d^I=12)\), see appendix~\ref{app:MLX}.

For the intermediate interaction strength, \(g=1.0\,\hbar\omega_B\ell_B\), already \(M=50\) PECs capture the qualitative overall behavior of the von Neumann entropy. However, the early-time dynamics after the quench is more demanding and consequently requires the inclusion of up to \(M=300\) PECs in order to achieve agreement with the ML-X reference. For the stronger interaction strength, \(g=2.0\,\hbar\omega_B\ell_B\), convergence becomes even more demanding. However, within the relevant time interval, the MCBO and ML-X results show the same qualitative behavior, with only small deviations remaining. Therefore, we conclude that the MCBO calculations are sufficiently converged and that the results presented in the main text are reliable.
\bibliography{citations}
\end{document}